\documentclass{ar-1col-S2O}
\usepackage{url}
\usepackage[numbers]{natbib}
\usepackage{graphicx}
\usepackage{subcaption}
\usepackage{bm}
\usepackage{slashed}
\usepackage{amssymb}
\usepackage{amsmath}

\newcommand{\slashpi}{\protect{\slash\hspace{-0.5em}\pi}}

\jname{Annu. Rev. Nucl. Part Sci.}
\jvol{AA}
\jyear{2024}
\begin{document}

% Page header
\markboth{Hayen}{Modern $\beta$ decay}

% Title
\title{Opportunities and open questions in modern $\beta$ decay}

%Authors, affiliations address.
\author{Leendert Hayen$^{1, 2}$
\affil{$^1$LPC Caen, ENSICAEN, Universit\'e de Caen, CNRS/IN2P3, Caen, France; email: hayen@lpccaen.in2p3.fr}
\affil{$^2$Department of Physics, North Carolina State University, Raleigh, 27607 North Carolina, USA}}

%Abstract
\begin{abstract}
For well over half a century, precision studies of neutron and nuclear $\beta$ decays have been at the forefront of searches for exotic electroweak physics. Recent advances in nuclear ab initio theory and the widespread use of effective field theories means that its modern understanding is going through a transitional phase. This has been propelled by current tensions in the global data set leading to renewed scrutiny of its theoretical ingredients. In parallel, a host of novel techniques and methods are being investigated that are able to sidestep many traditional systematic uncertainties and require a diverse palette of skills and collaboration with material science and condensed matter physics. We highlight the current opportunities and open questions with the aim of facilitating the transition to a more modern understanding of $\beta$ decay.
\end{abstract}

%Keywords, etc.
\maketitle

%Table of Contents
\tableofcontents

\section{INTRODUCTION}
\label{sec:introduction}
This year marks the 90th anniversary of Fermi's description of $\beta$ decay \cite{Fermi1934}, the experimental observation of $\beta^+$ decays \cite{Joliot-Curie1934}, and the prediction of the electron capture process \cite{Bethe1934}. Throughout its history, it has been at the foundation of the discovery of the neutrino, the violation of parity in the weak interaction, and the inception of the Standard Model as we know it today. Reflecting in 1959, C.S. Wu wrote \cite{Wu1959}
\begin{extract}
    As is well known, beta decay is full of surprises and subtleties. Its apparent perversities have threatened us not once but twice with the abandonment of some of our cherished conservation laws.
\end{extract}
% \begin{marginnote}[]
% \entry{Spontaneous symmetry breaking}{when equations of motion obey a symmetry, whereas the solution to those equations does not.}
% \end{marginnote}
It set the stage for chiral symmetry breaking in the decay of the pion,
propelling the ideas of spontaneous symmetry breaking and culminating in their manifestation in gauge theories in the Brout-Englert-Higgs (BEH) mechanism. In parallel, early studies of radiative corrections strengthened the idea of universality in which leptonic and hadronic weak decays are precipitated by the same underlying interaction. Subsequent calculations were sufficiently precise to estimate the mass of the BEH boson within a narrow range even before its experimental detection in 2012 \cite{Sirlin2013}.

Despite its tremendous success, the Standard Model of particle physics is known to be incomplete and falls short on a number of fundamental questions. Examples include the mass mechanism of neutrino's, the nature of dark matter, baryogenesis and, famously, the inclusion of gravity. Even so, as a renormalizable gauge theory built on a combination of relatively simple Lie groups, $SU(3)_c\times SU(2)_L\times U(1)_Y$, it yields impressive agreement up to extremely high energy scales (tens to hundreds of TeV) and precision (a part in $10^{13}$ for the magnetic moment of the electron). In the absence of strong indicators for specific extensions of the Standard Model, the current situation appears in some ways similar to that of over half a century ago, in which one leaves open the existence of effective operators constrained by Lorentz symmetry and searches for deviations from unitarity. The former may be recovered from the work for which Lee and Yang received the 1957 Nobel Prize \cite{Lee1956},
\begin{equation*}
\mathcal{H}_{\beta} = \sum_{i=V,A,S,T,P} \bar{p}O_i n \bar{e}O_i(C_i-C_i^\prime \gamma_5) \nu.
\end{equation*}
Even though the Standard Model is inherently \textit{vector minus axial-vector}\footnote{This was audaciously proposed by Sudarshan and Marshak in the face of contradicting experimental findings at a conference in Padua in 1957 \cite{Sudarshan1958}, noting experiments 'should be redone'.} \cite{Weinberg2009}, other interactions (Scalar, Tensor, and Pseudoscalar) are expected to show up in generic extensions. In this description, $C_i$ are generic couplings and necessarily of dimension [mass]$^{-2}$ pointing to the relevant energy scale. Standard Model interactions, for example, contain $\sqrt{2}C_V \approx 1\times 10^{-5}$ GeV$^{-2}$ so that $\sqrt{4\pi \alpha /2C_V} \approx 100$ GeV $\sim M_W$, while Wilson coefficients of exotic interactions are
\begin{eqnarray*}
    C_X &\propto& \left(\frac{M_W}{\Lambda_\mathrm{BSM}}\right)^{n \geq 2} \quad \mathrm{so~ that}  \\
    &\leq& 0.01\%\quad \mathrm{means} \quad \Lambda_\mathrm{BSM} \geq 10~\mathrm{TeV}
\end{eqnarray*}
\begin{marginnote}[]
\entry{Wilson coefficient}{A matching coefficient between a high and low energy theory for effective operators with only light fields.}
\end{marginnote}
appearing at some Beyond Standard Model scale $\Lambda_\mathrm{BSM}$. This immediately identifies the relevant scale for $\beta$ decays to be directly competitive with direct searches at colliders, as constraints (or discoveries) of $C_X$ at the 0.01\% level probe physics at the $\mathcal{O}(10)$ TeV scale.

\begin{figure}
    \centering
    \includegraphics[width=\textwidth]{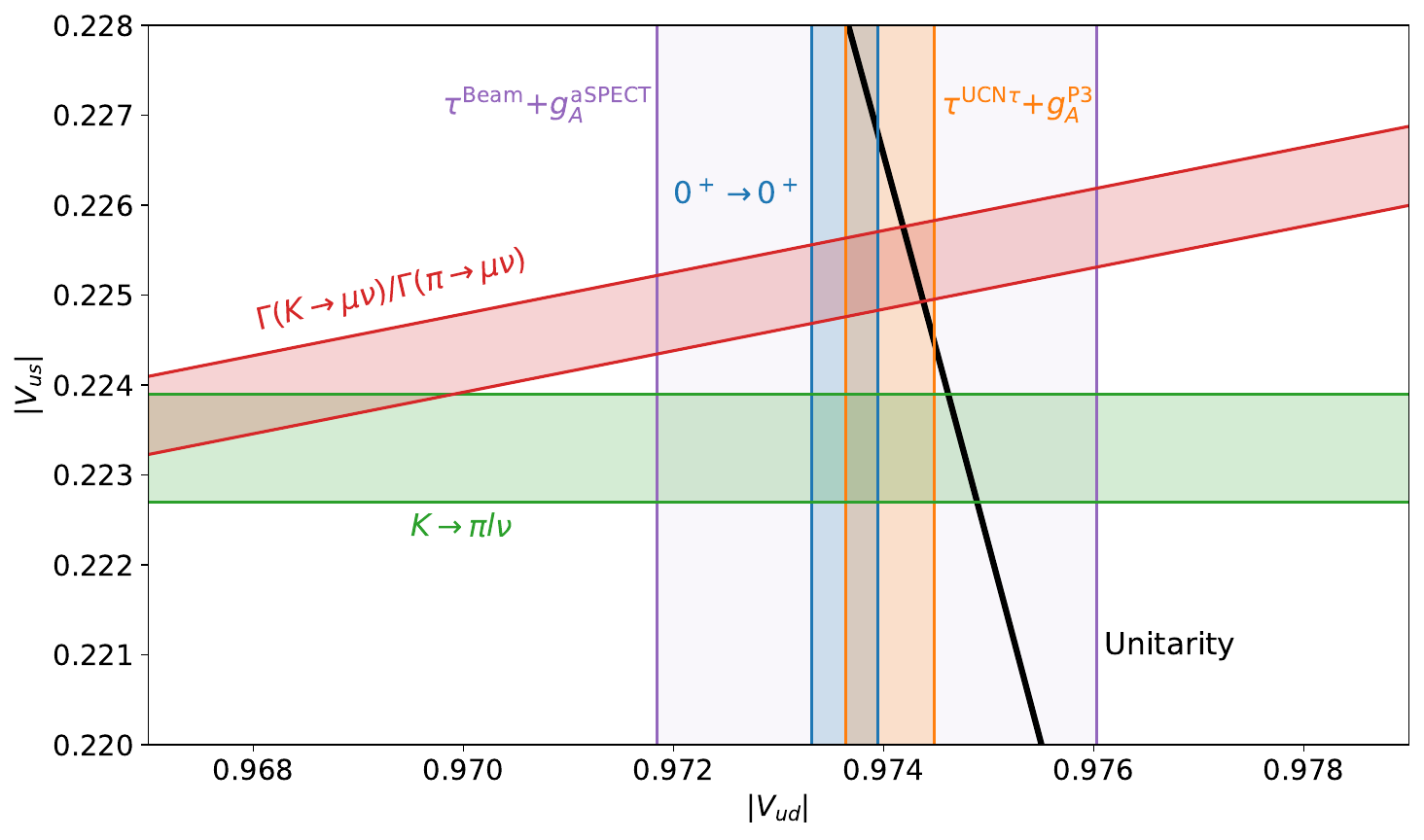}
    \caption{Selection of current results in the top row CKM unitarity test. The most precise determination of $V_{ud}$, obtained through $0^+\to 0^+$ nuclear decays, is not simultaneously compatible with either \textit{Kl2} or \textit{Kl3} determinations of $V_{us}$ and the unitarity circle. Additionally, the latter two disagree substantially, leading to the 'Cabibbo Angle Anomaly'. For the neutron the combination of the most precise individual measurements is compatible with \textit{Kl2} and unitarity. The same appears, somewhat surprisingly, for the combination of the `beam' neutron lifetime and determination and the aSPECT experiment, while both individually disagree with others at several standard deviations.}
    \label{fig:Vud_Vus_summary}
\end{figure}

The study of the Cabibbo-Kobayashi-Maskawa matrix, on the other hand, which relates weak and mass eigenstates by a $3\times 3$ unitary matrix
\begin{equation*}
    \left(\begin{array}{c}
                 d\\
                 s \\
                 b
            \end{array}\right)_\mathrm{w} =
        \left(\begin{array}{ccc}
             V_{ud} & V_{us} & V_{ub} \\
             V_{cd} & V_{cs} & V_{cb} \\
             V_{td} & V_{ts} & V_{tb}
        \end{array} \right) \left(\begin{array}{c}
            d \\
            s \\
            b
        \end{array} \right)_\mathrm{m}
    \end{equation*}
probes exotic physics at the TeV scale by testing the unitarity requirement of the top row at the 0.01\% level. More specifically, we may write
\begin{equation*}
    \Delta_\mathrm{CKM} = |V_{ud}|^2+|V_{us}|^2-1 \propto \left(\frac{M_W}{\Lambda_\mathrm{BSM}}\right)^2
\end{equation*}
where we neglected $|V_{ub}|^{2} \sim 10^{-5}$. Beta decay provides the largest matrix element of the top row, $V_{ud}$, and therefore provides the largest weight in terms of the total uncertainty in the unitarity sum. Current determinations of $\Delta_\mathrm{CKM}$ are inconsistent with unitarity at a few standard deviations, with different determinations showing also internal inconsistencies known as the `Cabibbo Angle Anomaly', with a selection of the data shown in Fig. \ref{fig:Vud_Vus_summary}.

%This has been in part spurred on by the use of dispersion relations in the calculation of electroweak radiative corrections, which showed substantially differing results due to an improved treatment of the non-perturbative QCD effects. 
%While initial studies focused on the neutron, developments within nuclear systems (which to date provide the most precise input to $V_{ud}$) are taking off and take advantage of recent nuclear \textit{ab initio} approaches.

%Within the context of nuclear $\beta$ decay efforts using nuclear \textit{ab initio} methods have necessarily focused on low mass systems, such as $^{6}$He and $^{10}$C. Both nuclear decays weigh heavily in the global data set in constraining exotic physics, although their theoretical requirements differ. In the case of $^6$He, one is interested in measuring the spectrum to high precision, as will be discussed in Sec. XXX, meaning higher-order matrix elements need to be controlled at the few-percent level. Recently, new calculations have been performed using the No Core Shell Model and Quantum Monte Carlo methods, showing XXXX. In the case of $^{10}$C, on the other hand, one is interested in a consistent description of the nuclear Green's function throughout a large momentum regime up to several hundreds of MeV for the precise calculation of radiative effects. This is still an open problem, as will be discussed below.

In recent years, a resurgence in theoretical activity has substantially changed the landscape even when compared to just five years ago. For single nucleons the last years have seen significant process in a number of parallel avenues, spurred on by conceptual and computational advances as well as tensions in the current global data set. Much of this work has focused on non-perturbative QCD corrections in some form. More specifically, lattice QCD (LQCD) calculations of nuclear charges have progressed significantly in precision, while non-perturbative QCD elements within $\mathcal{O}(\alpha)$ radiative corrections have benefited from using dispersion relations and exploratory connections to LQCD. More recently, significant attention has been given to elements of the calculation that have not been changed in decades, such as the traditional Fermi function arising from soft final state interactions and the correct matching between the various scales using renormalization group methods. As such, we will attempt to be more explicit than is usual to emphasize the running coupling when discussing the different formalisms.
\begin{marginnote}[]
\entry{Lattice QCD}{A non-perturbative method of calculation of low-energy QCD effects in discretized, Euclidian space-time.}
\end{marginnote}

In parallel, several experimental efforts have provided vastly improved results in both neutron and nuclear systems. Increasingly precise experimental results for the neutron lifetime and axial vector coupling result in a $V_{ud}$ extraction with a precision approaching that of superallowed Fermi decays without nuclear structure corrections. New technologies, such as Cyclotron Radiation Emission Spectroscopy (CRES) and the use of Superconducting Tunnel Junctions (STJ), show great promise from initial results and provide alternative measurement schemes, typically with drastically different sources of systematic uncertainty.

We attempt to report on the current status of the following questions:
\begin{enumerate}
    \item What systems and observables are most sensitive to current unknowns?
    \item How does (strong) isospin breaking affect $g_{V, A}$ at $\mu = m_e$?
    \item How should one consistently include final state interactions to 0.01\%-level precision?
    \item What inputs are needed for understanding nuclear structure effects on $\mathcal{O}(\alpha)$ radiative corrections?
\end{enumerate}

The paper is organized as follows: In Sec. \ref{sec:formalism} we introduce the different formalisms at each scale and describe the matching procedure. Section \ref{sec:available_experimental_observables} treats available experimental systems and observables and summarizes the current experimental status. Theoretical corrections to the nucleon and nuclear systems are described in Secs. \ref{sec:nucleon_level_corrections} and \ref{sec:nuclear_structure}, respectively. Current constraints on exotic physics are briefly summarized in Sec. \ref{sec:BSM_limits}, followed by a conclusion in Sec. \ref{sec:conclusions}. Finally, Appendix \ref{app:renormalization} attempts to provide a brief pedagogical introduction to renormalization group-improved perturbation theory while Appendix \ref{app:recoil_order} discusses recoil-order corrections.

\section{FORMALISM}
\label{sec:formalism}
Since the appearance of the Lee-Yang Hamiltonian in the 1950's, the description of $\beta$ decay has gone through several iterations with the development of the Standard Model and more recently using effective field theories. A decade ago, the introduction of the quark-level effective Lagrangian below the weak scale allowed one to directly compare constraints from collider searches to those of weak decays in a model-independent way for new UV physics \cite{Bhattacharya2012}. In parallel, increases in computational power have made \textit{ab initio} nuclear structure studies feasible for large portion of the nuclear chart, with effective interactions that are derived from Quantum Chromodynamics (QCD). In doing so, variables calculated through an expansion in a small parameter inherent to the separation of scales of effective field theories allow for quantifiable theoretical uncertainties, a process which is both difficult and fundamentally incomplete in traditional approaches.

One may construct a so-called \textit{EFT tower}, which connects interaction terms in the Lagrangians of different effective theories at their respective energy scale and degrees of freedom through a matching procedure. Short-distance phenomena in the high energy theory are `integrated out' order-by-order and are contained in the Wilson coefficients of effective operators using only the light degrees of freedom of the underlying theory, without inherent loss of information. Exotic interactions occurring at high scales may therefore be probed at each energy scale and results may be inter-compared following appropriate running and matching procedures.

\subsection{Quark-level Lagrangian}
\label{sec:quark_level_L}
Constructing an EFT using Standard Model fields applicable above the weak scale results in the SMEFT, which due to its particle content contains a staggering number of operators. Restricting to only dimension-six four-fermion operators most interesting for $\beta$ decay searches leaves 2499 terms to be determined and enter when comparing to collider searches. Here, we will focus on the Low-energy EFT (LEFT), applicable below the weak scale ($\mathcal{O}(M_W)$), obtained after integrating out the $W$ and $Z$ bosons, and all quarks except for the up, down, and strange flavours. At a scale of $\mu\simeq2$ GeV, the effective Lagrangian then consists of the $SU(2)_L\times U(1)_{em}$ gauge symmetries and effective four-fermion operators \cite{Gonzalez-Alonso2018},
\begin{equation}
    \mathcal{L} = \mathcal{L}_{ude\nu}^{eff} + \mathcal{L}_{QED} + \mathcal{L}_{QCD}
\end{equation}
We focus on the first term, which we may write as
\begin{eqnarray}
    \mathcal{L}_{ude\nu}^{eff} &=& -\frac{G_FV_{ud}}{\sqrt{2}}\left[(1+\epsilon_L)C_\beta^r(\mu)\bar{e}\gamma_\mu(1-\gamma_5)\nu_e \cdot \bar{u}\gamma^\mu(1-\gamma_5)d \right. \nonumber \\
    &+&\tilde{\epsilon}_L\bar{e}\gamma_\mu(1+\gamma_5)\nu_e \cdot \bar{u}\gamma^\mu(1-\gamma_5)d + \epsilon_R\bar{e}\gamma_\mu(1-\gamma_5)\nu_e \cdot \bar{u}\gamma^\mu(1+\gamma_5)d  \nonumber \\
    &+& \tilde{\epsilon}_R\bar{e}\gamma_\mu(1+\gamma_5)\nu_e\cdot \bar{u}\gamma^\mu(1+\gamma_5)d +\epsilon_T\bar{e}\sigma_{\mu\nu}(1-\gamma_5)\nu_e \cdot \bar{u}\sigma^{\mu\nu}(1-\gamma_5)d\nonumber \\
    &+& \tilde{\epsilon}_T\bar{e}\sigma_{\mu\nu}(1+\gamma_5)\nu_e \cdot \bar{u}\sigma^{\mu\nu}(1+\gamma_5)d + \epsilon_S \bar{e}(1-\gamma_5)\nu_e \cdot \bar{u}d + \tilde{\epsilon}_S\bar{e}(1+\gamma_5)\nu_e \cdot \bar{u}d \nonumber \\
    &-& \left.\epsilon_P \bar{e}(1-\gamma_5)\nu_e \cdot \bar{u}\gamma_5 d - \tilde{\epsilon}_P \bar{e}(1+\gamma_5)\nu_e \cdot \bar{u}\gamma_5 d + \ldots \right] + h.c.
\end{eqnarray}
\begin{marginnote}[]
    \entry{$\overline{\mathrm{MS}}$}{Modified minimal subtraction}
\end{marginnote}
where $\mu$ is the $\overline{\mathrm{MS}}$ renormalization scale, the tildes refer to right-handed neutrino fields and dots refer to subleading operators\footnote{We use $\gamma^5 = i\gamma^0\gamma^1\gamma^2\gamma^3$, $\sigma^{\mu\nu} = \frac{i}{2}[\gamma^\mu, \gamma^\nu]$ and set $\hbar=c=1$.}. The Fermi coupling constant is scale-independent and extracted from precise measurements of the muon lifetime \cite{Tishchenko2013},
\begin{eqnarray}
    G_F &=& \frac{\pi \alpha(\mu)g(\mu)}{\sqrt{2}M_W^2(\mu)s^2_W(\mu)} \\
    &=& \tau_\mu^{-1/2}\left[\frac{192\pi^3}{m_\mu^5}\left(1+\delta_\mu\right)\right]^{1/2} \nonumber
\end{eqnarray}
expressed in terms of Standard Model parameters ($s_W^2 = 1-M_W^2/M_Z^2$; $g(\mu) \approx 1$ is the running $SU(2)_L$ coupling at tree level), and $\delta_\mu$ are SM corrections. The $\epsilon, \tilde{\epsilon}$ are complex-valued coupling constants sensitive to new physics at some characteristic scale $\Lambda \gg M_W$, such that on dimensional grounds one has
\begin{equation}
    \epsilon_i, \tilde{\epsilon}_i \propto \left(\frac{m_W}{\Lambda}\right)^{n\geq 2}
\end{equation}
which are connected to SMEFT operators after appropriate running of the couplings using renormalization group methods. Similarly, $C_\beta^r(\mu)$ is the $\overline{\mathrm{MS}}$ Wilson coefficient which results from the matching of the SM and LEFT amplitudes and reads
\begin{eqnarray}
    C_\beta^r(\mu) = 1+\frac{\alpha}{\pi}\ln{\frac{M_Z}{\mu}} - \frac{\alpha\alpha_s}{4\pi^2}\ln{\frac{M_W}{\mu}} + \mathcal{O}(\alpha\alpha_s) + \mathcal{O}(\alpha^2)
    \label{eq:C_beta_r}
\end{eqnarray}
where $\alpha$ is the fine-structure constant and $\alpha_s$ is the strong interaction coupling. The renormalization group equation describing running of $C_\beta^r(\mu)$ was recently determined at the full two-loop level, discussed below.

As the interference between terms with right-handed neutrino's and the Standard Model tree-level amplitude is mitigated by the smallness of the neutrino mass (assuming $\mathcal{L}_{eff}^{m_\nu} = m_\nu \bar{\nu}_R\nu_L + h.c.$), it is common \cite{Gonzalez-Alonso2018} to neglect terms with $\tilde{\epsilon}$ to arrived at the 'linearized' low-energy effective Lagrangian
\begin{eqnarray}
    \mathcal{L}_\mathrm{eff} &=& -\frac{G_F\tilde{V}_{ud}}{\sqrt{2}}\left\{ C_\beta^r(\mu)\bar{e}\gamma_\mu(1-\gamma_5)\nu_e \cdot \bar{u}\gamma^\mu(1-(1-2\epsilon_R^\prime)\gamma_5)d + \epsilon_S \bar{e}(1-\gamma_5)\nu_e \cdot \bar{u}d\right. \nonumber \\
    &+& \epsilon_T\bar{u}\sigma_{\mu\nu}(1-\gamma_5)\nu_e \cdot \bar{u}\sigma^{\mu\nu}(1-\gamma_5)d - \epsilon_P\bar{e}(1-\gamma_5)\nu_e \cdot \bar{u}\gamma_5 d \} + h.c.
    \label{eq:linearized_LEFT}
\end{eqnarray}
leaving 9 couplings to be determined by experiment. The prime on $\epsilon_R^\prime$ is to denote the different running of the right-handed quark current compared to the left-handed one that goes like $C_\beta^r$. Differences are of order $\alpha \ln{M_Z/\mu}$ \cite{Dekens2019}, and become relevant in case of experimental observation of non-zero right-handed currents. 

Finally, one introduces $\tilde{V}_{ud}$ to take into additionally account potential changes affecting the muon lifetime but not neutron and nuclear $\beta$ decay,
\begin{equation}
    \tilde{V}_{ud} = V_{ud}(1+\epsilon_L+\epsilon_R^\prime)\left(1+\frac{\delta G_F}{G_F^\mu}\right),
    \label{eq:Vud_tilde}
\end{equation}
after dividing out common factors to directly show how CKM unitarity tests are sensitive to $\epsilon_{L, R}$ in this simplified scenario. Specifically, experimental tests of $\Delta_\mathrm{CKM}$ at low energy probe $\tilde{V}_{ud}$ and are sensitive to $\epsilon_{L, R}^{(\prime)}$ and $\delta G_F$.

\subsection{Nucleon-level Lagrangian}
\label{sec:nucleon_level_L}
\subsubsection{Lee-Yang Lagrangian}
\label{sec:Lee_Yang}
The nucleon-level Lagrangian was at its inception not thought of as an effective theory of some more UV-complete theory, but rather a linear combination of four-fermion operators obeying Lorentz invariance known as the Lee-Yang Lagrangian \cite{Lee1956}
\begin{equation}
\mathcal{H}_{\beta} = \sum_{i=V,A,S,T,P} \bar{p}O_i n \bar{e}O_i(C_i-C_i^\prime \gamma_5) \nu
\label{eq:Lee_Yang}
\end{equation}
where we may now interpret $C_i$ as Wilson coefficients evaluated at $\mu\sim m_e$. Using these equations, we may recover the nowadays well-known relationships between the Lee-Yang couplings and those of the LEFT, reproduced here for clarity:
\begin{equation}
    \begin{array}{rclcrcl}
    \bar{C}_V + \bar{C}_V^\prime & = & 2g_V\sqrt{1+\Delta_R^V}(1+\epsilon_L+\epsilon_R^\prime) & \qquad & \bar{C}_V - \bar{C}_V^\prime & = & 2g_V\sqrt{1+\Delta_R^V}(\tilde{\epsilon}_L+\tilde{\epsilon}_R) \\
    \bar{C}_A + \bar{C}_A^\prime & = & -2g_A\sqrt{1+\Delta_R^A}(1+\epsilon_L-\epsilon_R^\prime) & \qquad & \bar{C}_A - \bar{C}_A^\prime & = & 2g_A\sqrt{1+\Delta_R^A}(\tilde{\epsilon}_L-\tilde{\epsilon}_R) \\
    \bar{C}_S + \bar{C}_S^\prime & = & 2g_S\epsilon_S & \qquad & \bar{C}_S - \bar{C}_S^\prime & = & 2g_S\tilde{\epsilon}_S \\
    \bar{C}_P + \bar{C}_P^\prime & = & 2g_P\epsilon_P & \qquad & \bar{C}_P - \bar{C}_P^\prime & = & -2g_P\tilde{\epsilon}_P \\
    \bar{C}_T + \bar{C}_T^\prime & = & 8g_T\epsilon_T & \qquad & \bar{C}_T - \bar{C}_T^\prime & = & 8g_T\tilde{\epsilon}_T
    \end{array}
    \label{eq:C_relations}
\end{equation}
where we have extracted a common factor by defining $C_i = (G_FV_{ud}/\sqrt{2})\bar{C}_i$. At zero momentum transfer, the isovector charges $g_X$ are defined by the matrix element $g_X^{(0)}= \langle p| \bar{u} \Gamma_X d |n \rangle$ which may be calculated on the lattice discussed in Sec. \ref{sec:bare_charges_QCD}. The `inner' radiative correction, $\Delta_R^V$ is then the result of running the vector coupling down to the low energy scale \cite{Cirigliano2023a}, i.e.
\begin{equation}
    \Delta_R^{V,A} = [g_{V, A}(\mu)]^2\left(1+\frac{5\alpha(\mu)}{8\pi}\right)-1
\end{equation}
where $\mu$ is either equal to $m_N$ or $m_e$ depending on the approach, and the term in brackets results from agreement with the leading renormalization group evolution of $\delta_R^\prime$. 

\subsubsection{$\chi$PT and pion-less effective field theory}
\label{sec:chiPT}
In the previous discussion, the charges $g_X$ may be determined from lattice QCD and comparisons made against experimental results as a way of constraining new physics. Currently, however, lattice calculations are done assuming isospin symmetry, i.e. $m_u=m_d$ and similarly, $m_{\pi^0}=m_{\pi^\pm}$. In the absence of QED calculations with baryons on the lattice, an appropriate comparison with experiment requires the determination of all individual corrections as follows
\begin{equation}
    g_{V/A} = g_{V/A}^{(0)} \left[1+\sum_{n=2}^\infty \Delta_{V/A,\chi}^{(n)} + \frac{\alpha}{2\pi}\sum_{n=0}^\infty \Delta_{V/A, em}^{(n)} + \left(\frac{m_u-m_d}{\Lambda_\chi}\right)^{n_{V/A}}\sum_{n=0}^\infty\Delta_{V/A, \delta m}^{(n)}\right]
    \label{eq:g_V_A_CHIPT_expansion}
\end{equation}
where correction terms correspond to chiral corrections ($\chi$), electromagnetic and strong isospin symmetry breaking. The latter, due to $m_u\neq m_d$, is known to be small, whereas the other two are subjects of current study. 

The use of effective field theory and matching provides a systematic way in which to disentangle these various effects, and using only nucleons as degrees of freedom we may similarly introduce the pion-less EFT ($\slashpi$EFT) Lagrangian as follows \cite{Cirigliano2022a}
\begin{eqnarray}
    \mathcal{L}_{\slashpi} &=& -\frac{G_FV_{ud}}{\sqrt{2}}\left\{\bar{e}\gamma_\mu (1-\gamma_5)\nu_e \left[\bar{N}(g_Vv^\mu-2g_AS^\mu)\tau^+N\right.\right. \nonumber \\
    &+& \left.\frac{i}{2m_N}\bar{N}(v^\mu v^\nu-g^{\mu\nu}-2g_Av^\mu S^\nu)(\overleftarrow{\partial}-\overrightarrow{\partial})_\nu\tau^+N \right] \nonumber \\
    &+&\frac{ic_Tm_e}{m_N}\bar{N}(S^\mu v^\mu-S^\nu v^\mu)\tau^+N(\bar{e}\sigma_{\mu\nu}(1-\gamma_5)\nu) \nonumber \\
    &+&\left.\frac{i\mu_\mathrm{weak}}{m_N}\bar{N}[S^\mu, S^\nu]\tau^+N\partial_\nu(\bar{e}\gamma_\mu(1-\gamma_5)\nu)\right\}
\end{eqnarray}
with $N^T = (p, n)$ an isospin doublet of nucleons, $\tau^+$ is an isospin ladder operator while $v^\mu$ and $S^\mu$ represent the velocity and spin of the nucleon, respectively. By matching to an EFT in which pions are dynamical degrees of freedom (such as chiral perturbation theory, $\chi$PT), we may determine the effects of, e.g., the pion mass splitting as discussed in Sec. \ref{sec:gA_radiative_corrections}.

\subsection{Nuclear-level EFT}
\label{sec:nuclear_level_L}

In the spirit of the effective field theory one may add another story to the tower if there is an additional separation of scales for the process that one is interested in. Substantial amounts of effort have been focused on chiral EFT ($\chi$EFT), in which nucleons and pions are explicit degrees of freedom. Using the $\chi$EFT machinery, many-body currents may be constructed in an \textit{ab initio} fashion and used as inputs in many-body solution schemes. The last decades have seen tremendous progress in a variety of nuclear many body methods, such as the No Core Shell Model, Quantum Monte Carlo, Green's Function Monte Carlo, Coupled Cluster, In-Medium Similarity Group Renormalization, among others \cite{Hammer2020, Hergert2020, Tews2022}. One of the recent successes was the demonstration of the solution of the $g_A$ quenching problem \cite{Gysbers2019}, showing that by properly taking into account nucleon-nucleon correlations in the wave function and including many-body currents constructed using $\chi$EFT there is no need to \textit{a posteriori} rescale the Gamow-Teller strength to match experiment.

Historically, however, one made use of nuclear EFT formulations \textit{avant la lettre} through the form factor decomposition of the nuclear beta decay Hamiltonian. This has been performed either in a manifestly covariant fashion for allowed decays by Holstein \cite{Holstein1974}, or multipole decompositions by a variety of authors, notably Behrens and B\"uhring \cite{Behrens1982}, and Donnelly and Walecka \cite{Donnelly1975, Walecka2004}. The latter does not distinguish between the type of transition, but pays the price in breaking manifest Lorentz covariance by being able to perform a clean multipole decomposition only in the Breit frame. Nevertheless, both approaches are heavily used in experimental analyses and the calculation of total decay rates. Specifically, for a nuclear beta decay interaction term of the following form
\begin{equation}
    \mathcal{H}_\beta(0) = \frac{G_F}{\sqrt{2}}V_{ud}H_\mu(0)L^\mu(0)
    \label{eq:H_current_current}
\end{equation}
with $L^\mu = \bar{u}(p_e)\gamma^\mu(1-\gamma^5)v(p_\nu)$ one may, for example, perform a multipole decomposition of the time component of $H_0$ as follows
\begin{equation}
    \langle f e \nu| H_0 L^0 | i \rangle = \sum_{LM}\mathcal{C}_{m_im_f;M}^{J_iJ_f;L}Y^M_L(\hat{q})\frac{(qR)^L}{(2l+1)!!}F_L(q^2),
    \label{eq:H_0_decomp}
\end{equation}
where $\mathcal{C}$ contains a Wigner-$3j$ symbol, $q=p_f-p_i$, $Y_M^L$ is a spherical harmonic and $R$ is the nuclear radius so that $qR \ll 1$. Recently, new work has been performed to extend the multipole decomposition for scalar, pseudoscalar and tensor-type interactions \cite{Glick-Magid2023} and, separately, a systematic expansion of the multipoles into $\chi$EFT currents \cite{King2022}. For all $\beta$ decays other than that of the neutron, the calculation the vast majority of theoretical input is calculated in this type of formalism. Following results regarding potential double-counting or improper introduction of UV regulators in the Fermi function (Sec. \ref{sec:higher_order_QED}), calls for a consistent EFT analysis ring louder in the face of the tension of the current global data set.

\section{AVAILABLE EXPERIMENTAL OBSERVABLES}
\label{sec:available_experimental_observables}
Despite being sensitive only to the up-down quark sector in the electroweak Standard Model, the variety of probes and observables make (nuclear) $\beta$ decay a continuously compelling system to probe Beyond the Standard Model physics with a precision equal to or exceeding that or collider searches. 
%As is well known and summarized briefly in the following section, unitarity tests of the CKM matrix are one of the most precise, model-independent ways of constraining a variety of exotic new interactions and weigh heavily in full electroweak precision tests from low energy to collider searches. This is, in part, both due to the construction of an impressive data set in superallowed beta decays, as well as targeted effects using the neutron and selected mirror nuclei.

\subsection{Available experimental systems}
\label{sec:available_experimental_systems}
The choice of system is typically determined via the simplicity of its theoretical description (meaning the lowest degree of theoretical uncertainty due to, e.g., nuclear structure) and potential enhancements it provides in any of its observables. In the past, these have traditionally focused on \textit{allowed} Fermi or Gamow-Teller decays such as the neutron, superallowed (SA) $0^+\to 0^+$ Fermi transitions, mirror transitions within an isospin $T=1/2$ doublet, and pure Gamow-Teller transitions within an $T=1$ triplet. Much success has been garnered from integrated quantities such as the total decay rate and angular correlations where one can construct (super-)ratios of numbers of counts in a detector array to compensate several systematic effects to first order leading to improved precision. 

In the last decade, substantial focus has been diverted again to direct spectroscopy measurements due to linear rather than quadratic sensitivity to small exotic couplings in the Fierz interference term showing up as a $1/E_e$ distortion to the $\beta$ spectrum. On the other hand, following the success of ion and atom traps, precision recoil spectroscopy is once more becoming an interesting experimental system with the advent of novel technologies. In summary, searches have traditionally focused on
\begin{itemize}
    \item \textbf{Neutron decay}: As the simplest baryonic system, the neutron does not suffer from nuclear structure uncertainties and, following recent progress in lattice QCD, has its theoretical inputs determined to high precision. Experimentally, neutrons are trappable only when ultracold (i.e. kinetic energies on the order of tens of neV) which face challenging production cross sections, beam transport, and can only be produced at a limited number of facilities worldwide.
% \begin{marginnote}[]
% \entry{Superallowed}{Exceptionally fast allowed $\beta$ decays due to isospin symmetry.}
% \end{marginnote}
    \item \textbf{Superallowed $0^+\to 0^+$ decays}: The Wigner-Eckart theorem severely restricts the number of contributing matrix elements, while isospin symmetry largely determines those remaining. The consistency of a large global data set represents a significant multi-decade effort of the community, but its theoretical corrections are currently in flux which limit its precision on $V_{ud}$. Resolving these is a high priority and requires significant investment from nuclear \textit{ab initio} theory. 
    %Constraints on exotic scalar currents are determined largely by its lowest mass decays ($^{10}$C, $^{14}$O) which are experimentally extremely challenging.
    \item \textbf{Mirror nuclei}: Mixed Fermi and Gamow-Teller decays allow for cancellations in angular correlations, thereby providing substantial (up to an order of magnitude for $^{19}$Ne) sensitivity gains on determining the mixing ratio, itself an ingredient in their $V_{ud}$ extraction. Isospin symmetry largely determines theoretical corrections to a satisfactory level, and its mixed decay allows for studies of CP-violation through, e.g., $D$-correlation measurements, and sensitivity to right-handed exotic currents.
\begin{marginnote}[]
\entry{Mirror transition}{Decay between $N=Z\pm 1$ nuclei within the same isospin $T=1/2$ doublet.}
\end{marginnote}
    \item \textbf{$T=1$ Gamow-Teller decays}: Pure Gamow-Teller decays within an isospin triplet share favourable traits of the categories above, with isospin symmetry resulting in a large Gamow-Teller matrix element and determining some recoil-order matrix elements, leading to a small theoretical uncertainty. 
    %Historically, much attention has been devoted to $^6$He and $^8$B for their high production yields, compatibility with ion traps and manageable theoretical calculations. 
    It is sensitive only to exotic tensor currents, however.
\end{itemize}
All of these categories describe \textit{allowed} decays, i.e. the lepton fields do not carry orbital angular momentum and leading-order matrix elements are large, with only sporadic efforts into, e.g., pseudoscalar ($\Delta J^\pi = 0^-$) \cite{Bhalla1960a} or unique forbidden decays. Many of the latter are of current interest in the prediction of reactor antineutrino fluxes, which form an essential ingredient in the analysis of the reactor antineutrino anomaly \cite{Hayes2014, Hayes2016, Hayen2019, Hayen2019b} and predictions for low-threshold coherent elastic neutrino scattering \cite{HayenIP}. Total absorption spectrometers \cite{Fijakowska2017}, offer interesting possibilities of sidestepping typical systematic effects in precision $\beta$ spectroscopy methods, such as bremsstrahlung and scattering losses. Recently, unique forbidden $\beta$ decays were also put forward as an interesting system for BSM searches \cite{Glick-Magid2016a} as its $\beta$ spectrum shape is sensitive also to right-handed currents.

\subsection{$\beta$ spectrum shapes}
\label{sec:spectral_shapes}
Following integration over all but the $\beta$ particle energy, the differential decay rate may be written as follows \cite{Hayen2018},
\begin{equation}
    \frac{d\Gamma}{dW} = \frac{G_F^2\tilde{V}_{ud}^2}{2\pi^3}F(Z, W)C(Z, W) K(Z, W) pW(W_0-W)^2 \times \left(1+b_F\frac{1}{W}\right)
    \label{eq:spectrum_shape_general}
\end{equation}
where $W = E_e/(m_ec^2)$ ($p=\sqrt{W^2-1}$) is the electron energy (momentum) in dimensionless units, $F(Z, W)$ is the traditional Fermi function (see Sec. \ref{sec:higher_order_QED}), $C(Z, W)$ the shape factor containing most nuclear structure corrections, and $K(Z, W)$ higher-order corrections. Spectrum shape measurements typically do not control the overall normalization of the source activity and system efficiencies to a level where one is sensitive to $\tilde{V}_{ud}$. Instead, sensitivity to BSM occurs through the determination of the Fierz interference term,
\begin{equation}
    b_F = \pm 2 \gamma \frac{1}{1+\rho^2}\mathrm{Re}\left\{\frac{g_S\epsilon_S}{g_V(1+\epsilon_L+\epsilon_R^\prime)}+ \rho^2 \frac{4g_T\epsilon_T}{-g_A(1+\epsilon_L - \epsilon_R^\prime)} \right\},
    \label{eq:bF}
\end{equation}
where $\gamma = \sqrt{1-(\alpha Z)^2}$. Its linearity in exotic couplings and simple kinematic signature means it typically forms the backbone of most spectroscopy-related $\beta$ BSM searches. In fact, a single direct measurement of the spectrum shape could suffice to obtain competitive precision, which has reinvigorated direct spectrum shape measurements in the last decade \cite{Severijns2008, Naviliat-Cuncic2009}. 

The theoretical corrections entering Eq. (\ref{eq:spectrum_shape_general}) were recently reevaluated and extended to include analytical descriptions reaching a precision of down to a part in $10^4$ \cite{Hayen2018, Hayen2019a}. Further efforts were performed in determining and contextualizing the magnitude of recoil-order and radiative corrections \cite{Vanlangendonck2022} and nuclear ab initio calculations for $^6$He were performed using Quantum Monte Carlo \cite{King2022} and No Core Shell Model \cite{Glick-Magid2022} methods. The multipole decomposition methods used for the vector and axial currents were extended to include scalar and tensor interactions \cite{Glick-Magid2023}.

The dominant recoil-order matrix element to contribute for allowed decays is the so-called weak magnetism contribution (see App. \ref{app:recoil_order}), which has recently been discussed in the largest compilation to date \cite{Severijns2023}. Its evaluation is of paramount importance for spectral shape and correlation measurements of both $T=1/2$ mirror and $T=1$ pure Gamow-Teller transitions, but also contributes significantly to the reactor antineutrino anomaly analysis is the $80 < A < 140$ mass range. For the latter, some recent efforts have made inroads \cite{DeKeukeleere2024}.

The statistical sensitivity to a Fierz-like term in a spectrum shape measurement was investigated a number of years ago \cite{Gonzalez-Alonso2016}, where one found an optimal endpoint on the order of 2 MeV. This can intuitively be understood by the strong correlation between the overall spectrum normalization and Fierz term for low endpoints (where $m_e/E_e \sim 1$) and a diluted effect of the $1/E_e$ term at high energies and endpoints. Spectral measurements are extremely sensitive to calibration non-linearities, however, \cite{Hayen2020a}, which often turn out to be the dominant uncertainty in scenarios where backscattering can be mitigated. Lower-endpoint decays can then, for example, benefit from more calibration points to constrain non-linearities and compensate in statistics through longer running times or higher count rates.

\subsubsection{Experimental status}
A comprehensive overview of $\beta$ spectrum shape efforts was provided in Ref. \cite{Gonzalez-Alonso2018}. Measurements of the $\beta$ spectrum shapes of superallowed decays have been few in the past \cite{George2014} due to the limited sensitivity one may hope to attain compared to integrated decay rates (see below). With the appearance of additional energy-dependent radiative corrections, discussed in Sec. \ref{sec:delta_NS}, such a measurement could be highly valuable.

Instead, experiments have focused on mirror and pure Gamow-Teller transitions. At NSCL, measurements of the $\beta$ spectrum shape of $^6$He were performed and is currently under analysis \cite{Huyan2016, Huyan2019}. Similar efforts are underway at GANIL with the b-STILED experiment \cite{Kanafani2023}.

Following the importance of forbidden $\beta$ decays \cite{Hayes2014, Hayen2019, Hayen2019b} in the prediction of the reactor antineutrino flux and its implications for the reactor anomaly and spectral disagreement \cite{Hayes2016}, several measurements campaigns have started for transitions within the fission fragment region \cite{Fallot2019} using total absorption spectroscopy techniques \cite{Fijakowska2017, Rice2017}.

A novel technique, called Cyclotron Radiation Emission Spectroscopy \cite{Asner2015a}, was used for the first time in the broadband measurements of a continuous $\beta$ decay \cite{Byron2023a}. Here, a ratio between spectra of $^6$He and $^{19}$Ne was taken, showing a proof of principle in which several systematic uncertainties cancel to first order and receive an enhanced sensitivity to a Fierz term due to its sign change for $\beta^\pm$ decays.

Finally, using the energy dependence of the angular correlations measured in neutron decay, several experiments were able to place limits on a Fierz term \cite{Sun2020, Beck2023, Saul2020}, where both aSPECT and PERKEO III reach a precision on the order of 2\%.

\subsection{Integrated decay rates}
\label{sec:integrated_decay_rates}
\begin{figure}[ht]
    \centering
    \includegraphics[width=\textwidth]{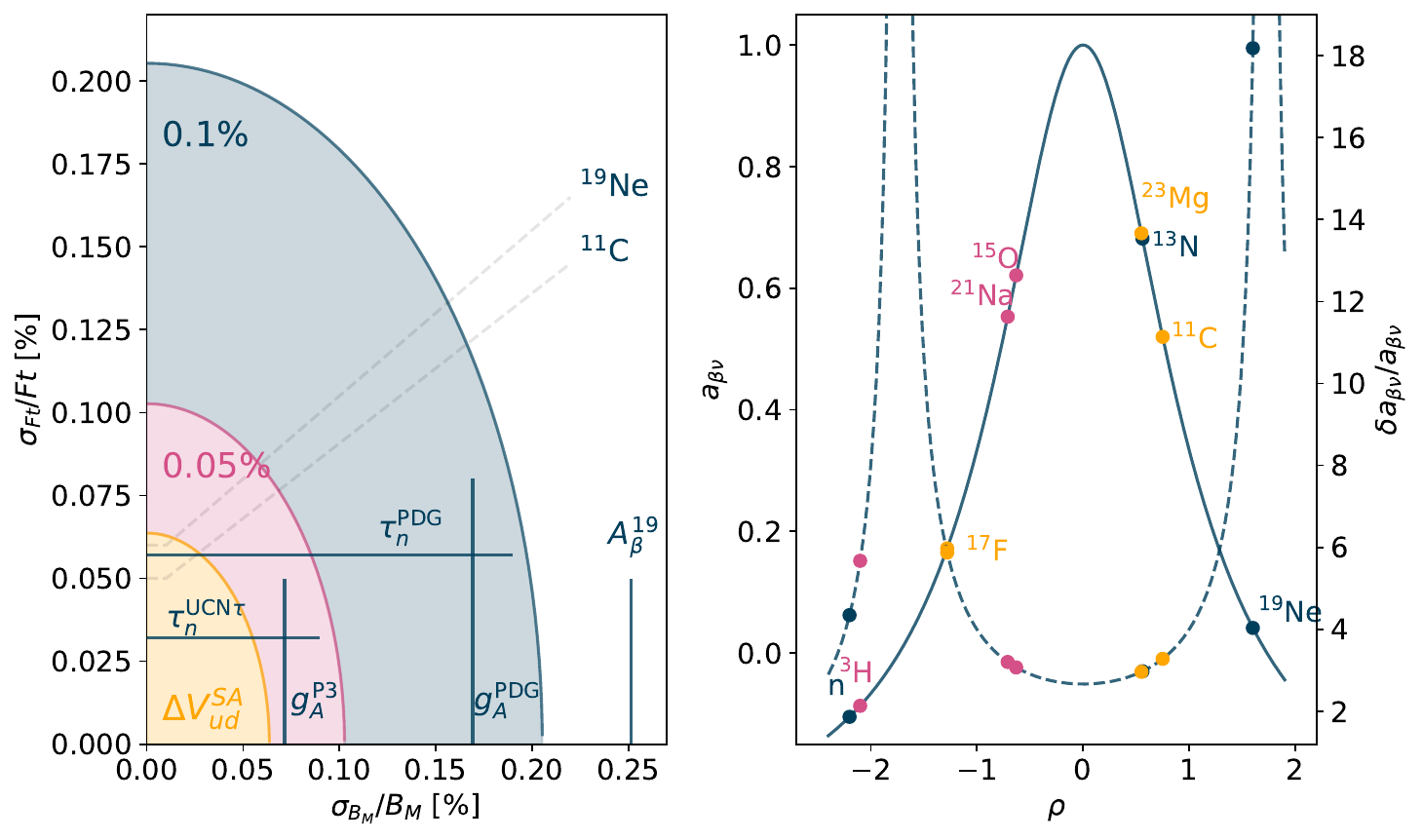}
    \caption{Summary of the $V_{ud}$ sensitivity and current status for mixed decays. (Left) Relative uncertainties of $\mathcal{F}t$ versus those of $B_M = 1+(f_A/f_V)\rho^2$ together with curves of constant $\Delta V_{ud}$. For the neutron we show both PDG averages for $g_A$ and $\tau_n$ together with the most precise individual measurements (UCN$\tau$ for $\tau_n$ \cite{Gonzalez2021} and PERKEO3 for $g_A$ \cite{Markisch2019}). Shown also is the most precise mirror determination for $\rho$ \cite{Combs2020}, and $\mathcal{F}t$ values below 0.1\% and $A < 20$ \cite{Severijns2023}. (Right) Values for $a_{\beta\nu}$ for the lowest mass mirror nuclei, together with the enhancement factor between $a_{\beta\nu}$ and $V_{ud}$. The latter is to be interpreted as follows: a relative determination of $\delta a_{\beta\nu}/a_{\beta\nu} = 1$\% for $^{19}$Ne results in an uncertainty 18 times smaller, i.e. $\delta V_{ud} = 0.055$\% with the uncertainty solely from $\delta a_{\beta\nu}$.}
    \label{fig:sensitivity_mirrors_summary}
\end{figure}

We may schematically write the decay rate for a semi-leptonic weak decay as follows
\begin{equation}
    \Gamma = G_F^2~ |V_{rs}|^2~(1+\mathrm{RC})~|\langle \mathcal{O}_{\mathrm{hadr}}\rangle |^2 \times (\mathrm{phase~space})
    \label{eq:Gamma_semileptonic_schematic}
\end{equation}
where $V_{rs}$ it the $\{r, s\}$ element of the CKM matrix, RC are radiative corrections and $\mathcal{O}_{\mathrm{hadr}}$ represents the hadronic matrix element(s). Besides $G_F$, a determination of $V_{rs}$ is then possible using a $\beta$ decay after the measurement of such quantities: ($i$) the $Q$ value, as input to the phase space; ($ii$) the half-life; ($iii$) the branching ratio; ($iv$) the hadronic matrix element, if it is not given by underling symmetries. For allowed decays we may write Eq. (\ref{eq:Gamma_semileptonic_schematic}) in the conventional fashion
\begin{equation}
f_Vt\left[1+\frac{f_A}{f_V}\rho^2\right](1+\delta_R^\prime)(1-\delta_C+\delta_{NS}) = \frac{K}{G_F^2\tilde{V}_{ud}^2(1+\Delta_R^V)|M_F^0|^2}
\label{eq:ft_mixed}
\end{equation}
where $f_{V, A}$ is the integral over the $\beta$ spectrum of Eq. (\ref{eq:spectrum_shape_general}), $\rho$ is the Gamow-Teller to Fermi mixing ratio, discussed further in Sec. \ref{sec:mirror_delta_NS}, $\delta_{R,C,NS}$ are radiative, isospin-breaking, and nuclear-structure corrections, respectively, and $K/(\hbar c)^6 = 8120.276 236(12) \times 10^{-10}$ GeV$^{-4}$s. For superallowed Fermi decays $\rho=0$ while $M_F^0 = \sqrt{2}$, whereas mirror decays imply $M_F^0=1$ and $\rho$ may be determined from a measurement of an angular correlation (see Sec. \ref{sec:correlation_measurements}). While an equivalent expression may be written for pure Gamow-Teller decays, the leading matrix element is not determined by underlying symmetries and nuclear theory is currently unable to calculate it with sufficient precision (compare the $\sim 1\%$ precision from recent efforts on $^6$He \cite{Glick-Magid2022, King2022} to the few parts in $10^4$ from the SA and neutron data set). For all $\beta$ decays except the neutron, the bulk of Eq. (\ref{eq:spectrum_shape_general}) is calculated exclusively using the traditional, multipole-decomposition methods discussed in Sec. \ref{sec:nuclear_level_L}, and forms the backbone of the $V_{ud}$ extraction discussed above through the calculation of $f_{V, A}$. Figure \ref{fig:sensitivity_mirrors_summary} shows a current overview of the precision on $V_{ud}$ from superallowed and mixed decays.

Substantial progress has been made on the theoretical front, with a dispersive approach allowing a substantially reduced uncertainty on $\Delta_R^V$ and connections to lattice QCD. Applying similar methods to nuclei unearthed an additional correction to $\delta_{NS}$, the evaluation of which is still in flux. Finally, efforts are underway for alternative determination of $\delta_C$. The uncertainties on these theoretical corrections determine the current uncertainty estimate on the global $V_{ud}$ determination, and will be discussed in Sec. \ref{sec:nuclear_structure}.

The integral decay rate retains a sensitivity to the Fierz term through the average value of $\langle m_e/E_e \rangle$, which ranges from 0.619 for $^{10}$C to 0.125 for $^{74}$Rb, as the endpoint increases for increasing mass \cite{Gonzalez-Alonso2018}.

\subsubsection{Experimental status} The nuclear data set used for the $V_{ud}$ extraction of superallowed Fermi decays comprises more than 200 individual measurements, and has been reviewed last by Towner and Hardy \cite{Hardy2020}. Generally, branching ratio measurements of $T=1$ superallowed decays are the limiting factor and are a significant priority for the community. In terms of physics reach, the current determination of scalar current limits is influenced most by the branching ratio uncertainty in the superallowed branch of $^{10}$C. A recent determination was performed at ISOLDE \cite{Blank2020}, with a number of experiments still in the planning stage.

For mirror decays, a large number of measurements in the past decade has brought significant improvements in the precision of almost all relevant transitions. These were recently reviewed by \cite{Severijns2023}, and show several isotopes with $ft$ values with a relative precision of a few parts in $10^4$. Low mass isotopes are additionally promising to make contact with nuclear \textit{ab initio} theory efforts, with $^{11}$C boasting a relative precision of 0.05\% on its $ft$ value.

The most precise determination of the neutron lifetime was obtained by the UCN$\tau$ collaboration \cite{Gonzalez2021, Cude-Woods2022}. Besides tests of the weak interaction \cite{Dubbers2021}, recent results using ultracold neutrons were additionally used to search for axion-like particles \cite{Ayres2023}, mirror-neutrons \cite{Abel2021, Broussard2022}, and dark matter candidates \cite{Klopf2019, Sun2018a} (see Sec. \ref{sec:BSM_limits}).

Finally, new measurements of the $^6$He \cite{Kanafani2022} and $^{20}$F \cite{Hughes2018} lifetimes were performed, resolving previous discrepancies.

\subsection{Angular correlations and recoil spectroscopy}
\label{sec:correlation_measurements}
At leading order, angular correlation measurements are sensitive to quadratic exotic couplings, although they receive sensitivity to left-handed new physics through an averaged Fierz interference term. Higher-order corrections terms were recently evaluated \cite{Hayen2020a}, and one may write the differential cross section after summing over final state helicities as
\begin{eqnarray}
    d\Gamma &=& \frac{G_F^2}{2\pi^3}F_0L_0K(W_e, W_0) p_eW_e(W_0-W_e)^2
    \left[f_0 +  \sum_{k \geq 1}f^{\beta\nu}_k P_k(\cos \theta_{\beta \nu}) + G_k(J_i)\biggl\{ f^{\sigma e}_k P_k(\cos \theta_e)\right. \nonumber \\
    &+& f^{\sigma \nu}_k P_k(\cos \theta_\nu) + f_k^{\sigma \times} P_k(\cos \theta_\times) \biggr\} + \mathrm{higher~orders~}\Biggr]dW_e d\Omega_e d\Omega_\nu,
    \label{eq:general_decay_rate}
\end{eqnarray}
where $G_k$ is a polarization tensor of rank $k$ of the initial state, $P_k(\cos \theta)$ a Legendre polynomial of degree $k$, $f_0$ is the isotropic shape factor of the previous section and $K(W_e, W_0)$ common corrections \cite{Hayen2018}. The `higher orders' in Eq. (\ref{eq:general_decay_rate}) stands for correlations involving more exotic combinations of momenta and higher powers (see, e.g., Ref. \cite{Ebel1957, Ivanov2013, Gudkov2006}), which we neglect here. The angles are defined as follows
\begin{equation}
    \cos \theta_{\beta\nu} = \frac{\vec{p}_e\cdot \vec{p}_\nu}{|\vec{p}_e||\vec{p}_\nu|}, \quad \cos \theta_e = \frac{\hat{J}\cdot \vec{p}_e}{|\vec{p}_e|}, \quad \cos \theta_\nu = \frac{\hat{J}\cdot \vec{p}_\nu}{|\vec{p}_\nu|}, \quad \cos \theta_\times = \frac{\hat{J}\cdot (\vec{p}_e \times \vec{p}_\nu)}{|\vec{p}_e||\vec{p}_\nu|},
\end{equation}
with $\hat{J}$ the initial nuclear polarization and neglecting final-state interactions (reduced by a factor $\alpha Z$), all $f_{k}^X$ with $k\geq 1$ depend only quadratically on exotic couplings. This implies that, contrary to allowed spectrum shape measurements, angular correlations are sensitive also to right-handed currents. Tensions in the global neutron data set, for example, show a preference for right-handed tensor currents as discussed in Sec. \ref{sec:BSM_limits}.

For mixed Fermi and Gamow-Teller decays, opposite contributions to various angular correlations leads to significant sensitivity enhancements to $\rho$, the Gamow-Teller to Fermi mixing ratio. The latter is a crucial input for $V_{ud}$ extractions (see Eq. (\ref{eq:ft_mixed})), and enhancements of over to an order of magnitude are possible in cases such as $^{19}$Ne. These have recently been reviewed \cite{Hayen2020a, Vanlangendonck2022}, and following significant community efforts in providing additional experimental inputs \cite{Severijns2008, Severijns2023}, several isotopes can contribute significantly to the global $V_{ud}$ extraction following an improved determination of $\rho$.

\subsubsection{Experimental status}
Angular correlation studies have focused on constructing experimental asymmetries using mirror (including the neutron), and pure Gamow-Teller transitions. Most of these focus on measurements of the $\beta$-$\nu$ angular correlation, $a_{\beta\nu}$, and the $\beta$-asymmetry, $A_\beta$. A number of experiments have published their final results in the last few years, including the UNCA \cite{Brown2018, Plaster2019}, PERKEO III \cite{Markisch2019}, aSPECT \cite{Beck2020, Beck2023} and aCORN \cite{Hassan2021} experiments. Currently, the Nab experiment at Oak Ridge National Laboratory \cite{Pocanic2009, Broussard2019} is on line and performing commissioning measurements. In the mean time, the preparation for the PERC facility \cite{Dubbers2008} is in progress and will enable more high-precision measurements of free neutron decay.

Significant efforts have been reported in pure Gamow-Teller $T=1$ decays, with a substantial improvement of the Beta Paul Trap result in the $A=8$ system \cite{Gallant2023, Burkey2022} enabled by dedicated theoretical efforts \cite{Sargsyan2022}. Similarly, new results were reported for the decay of $^6$He \cite{Muller2022} using a magneto-optical trap. A proof of principle for the $^{32}$Ar decay in the WISArD setup was shown to have good sensitivity \cite{Araujo-Escalona2020} with promises to reach sub-percent precision.

In mirror systems, a new analysis of the $\beta$-asymmetry measurement of $^{19}$Ne was reported \cite{Combs2020} resulting in the most precise $V_{ud}$ determination of a mirror isotope. In parallel with multiple theoretical efforts describing sensitivities and corrections \cite{Hayen2020a, Vanlangendonck2022}, the St. Benedict Facility at the University of Notre Dame aims to measure the mixing ratio of several mirror isotopes in the coming years \cite{Porter2023, Burdette2019, OMalley2020}.

Finally, the recoil energy spectrum alone is sensitive also to the $\beta$-$\bar{\nu}$ angular correlation, $a_{\beta\nu}$, without needing to detect multiple final states. Because of the strong imbalance in masses, changes in $a_{\beta\nu}$ cause $\mathcal{O}(1)$ distortions to the energy spectrum. This is the objective of the SALER experiment at FRIB using novel Superconducting Tunnel Junction detectors, which boast excellent linearity and full energy deposition after implantation. This will build upon the results obtained using $^7$Be electron capture studies with the BeEST experiment \cite{Fretwell2020, Friedrich2021}.

\section{NUCLEON-LEVEL CORRECTIONS}
\label{sec:nucleon_level_corrections}

Following tensions in the global data set and the onset of new methods in the calculation of radiative and nuclear structure corrections, many of the theoretical inputs to the $V_{ud}$ determination are being reviewed. While alternative calculations of isospin breaking corrections in nuclei have been investigated in parallel for several decades, the dispersive approach to an element of the radiative corrections in 2018 \cite{Seng2018, Seng2019b}, the so-called `inner' radiative correction $\Delta_R^V$, have accelerated the pace. Since the signs of CKM top-row non-unitarity caused by the 2.5$\sigma$ shift of the latter, progress has been made on isospin breaking corrections to the vector charge, first calculations of corrections to the axial charge, and EFT methods for its overall running and $\mathcal{O}(\alpha)$ behaviour. Finally, lattice QCD has reached percent-level determinations of the axial charge inviting comparison to experiment as a constraint on new physics. In this section, we will review and reinterpret a number of the results in the literature. 
 
\subsection{Lattice QCD and bare isovector charges}
\label{sec:bare_charges_QCD}
In making the connection between the quark-level Lagrangian of Eq. (\ref{eq:linearized_LEFT}) and the Lee-Yang Lagrangian of Eqs. (\ref{eq:Lee_Yang}) \& (\ref{eq:C_relations}), one introduced the nucleon isovector charges for the different types of fermion bilinear operators, $g_X = \langle p | \bar{u} \Gamma_X d | n \rangle$. Scalar and tensor couplings are zero at leading order in the Standard Model, meaning determinations of $g_{S,T}$ need to be determined only at the 10\% level for constraints on new physics to reach the multi-TeV level. The axial charge, on the other hand, is determined experimentally at the 0.1\% level (see also Sec. \ref{sec:correlation_measurements}), and comparisons with \textit{ab initio} determinations at a similar precision level allow for stringent constraints on right-handed exotic interactions (see Secs. \ref{sec:gA_radiative_corrections} and \ref{sec:BSM_limits}). The vector and pseudoscalar charges can be determined from underlying Standard Model symmetries, using the Conserved Vector Current (CVC) and Partially Conserved Axial Current (PCAC), respectively. The vector charge, $g_V$, receives corrections only at second order in isospin symmetry breaking effects (see below), and is not calculated directly in isospin symmetric lattice QCD calculations. The pseudoscalar charge, $g_P$, while substantially greater than unity, does not significantly contribute to the cross section due to the small momentum transfer in beta decay compared to initial and final state masses \cite{Gonzalez-Alonso2014}.

The remaining three charges ($g_{S, T, A}$) may be calculated on the lattice through direct calculation of two- and three-point correlators and appropriate current insertions. Unlike perturbative calculations where one renormalizes the loop-level results by introducing a UV renormalization scale $\mu$ in, e.g., the $\overline{\mathrm{MS}}$ scheme (see App. \ref{app:renormalization}), the inverse lattice spacing $1/a$ acts as a UV regulator\footnote{Note that no momenta greater than $2\pi/a$ may exist in discretized space-time. While procedures exist to transform results in an equivalent $\overline{\mathrm{MS}}$ scheme, neither vector nor axial vector charges run in pure QCD (see Eq. (\ref{eq:C_beta_r})).}. Its total uncertainty is determined by statistical effects from calculating the path integral using Monte Carlo methods, as well as systematic uncertainties following extrapolation to the continuum (i.e. vanishing lattice spacing), infinite volume, and setting the pion at its physical mass. The latter is of relevance as for sources and sinks separated by a time $\tau$, the signal-to-noise ratio decreases like $\exp[-(m_N-3m_\pi/2)\tau]$, with $m_{N(\pi)}$ the nucleon (pion) mass. Typically, calculations do not exceed $\tau \sim 1.5$ fm, and many older calculations were performed at much heavier pion masses due to exponentially suppressed computational requirements. While extrapolations to the physical pion mass can be made using chiral perturbation theory, their convergence properties are unclear and modern FLAG summaries consider only calculations with at least one point at the physical pion mass for their global averages \cite{Aoki2021}.

In 2005, the first LQCD calculations of $g_A$ with relatively light dynamical quarks was performed \cite{Edwards2006}, in agreement with the experimental data after extrapolation to the physical pion mass with a 7\% relative uncertainty. Progress was therefore expected to be forthcoming, but instead the community was faced with confounding results for the next decade. Excited state contaminations (ESC) were found to be a major culprit and remain the predominant concern to this day \cite{Aoki2021, Jang2020, He2021, Meyer2022, Nicholson2021}. This can be understood from a spectral decomposition of the three-point function \cite{He2021},
\begin{eqnarray}
    C_\Gamma(\tau;t_\mathrm{sep}) &=& \sum_{\bm{y}, \bm{x}} \langle \Omega | N(t_{\mathrm{sep}}, \bm{y})j_\Gamma(\tau, \bm{x})N^\dagger(0, \bm{0}) | \Omega \rangle \\
    &=& \sum_n |z_n|^2g_{nn}^\Gamma e^{-E_nt_{\mathrm{sep}}} + 2 \sum_{n< m} z_n z_m^\dagger g_{nm}^\Gamma e^{-(E_n+\frac{\Delta_{nm}}{2})t_{\mathrm{sep}}} \cosh\left[\Delta_{nm}(\tau-\frac{t_{\mathrm{sep}}}{2})\right] \nonumber
\end{eqnarray}
where $t_{\mathrm{sep}}$ is the separation between the source and sink, $\tau$ denotes the current insertion time, $z_n = \langle \Omega | N | n \rangle$ for the creation ($N^\dagger$) or annihilation ($N$) of states with quantum numbers of the nucleon from the vacuum ($|\Omega \rangle$), $\Delta_{nm} = E_n-E_m$ and $g_{nm} = \langle n | j_\Gamma | m \rangle$ are the individual matrix elements. Excited states are plentiful, consisting of (radial excitations of) $N(\bm{p})\pi(-\bm{p})$, $N(\bm{0})\pi(\bm{0})\pi(\bm{0})$, $\ldots$ towers which become more problematic for lower pion masses. The calculation of the axial charge, in particular, contains large corrections due to ESC because of the large coupling of the axial current to the pion field ($z_n$). A number of methods have been proposed to improve the precision, with two calculations published with a sub-percent precision \cite{Chang2018, Walker-Loud2020}. Efforts have been proposed to reach an 0.5\% precision with current computer architecture and down to 0.2\% in the exascale computing era. 

\begin{table}[ht]
\tabcolsep7.5pt
\caption{Overview of the current Flavour Lattice Averaging Group results for the various nucleon isovector charges.}
\label{tab:nucleon_charges}
\begin{center}
\begin{tabular}{@{}l|c|c|c|c|c@{}}
\hline
{M}ethod & $g_V$ & $g_A$ & $g_S$ & $g_T$ & $g_P$\\
\hline
FLAG 21 ($N_F=2+1$) & - & 1.248(23) & 1.13(14) & 0.965(61) & -\\
FLAG 21 ($N_F=2+1+1$) & - & 1.246(28) & 1.02(10) & 0.989(34) & - \\
CVC/PCAC & 1 & - & 1.02(10) & - & -349(9)\\
\hline
\end{tabular}
\end{center}
\label{tab:nucleon_charges}
\end{table}

% \begin{table}
%     \centering
%     \caption{Overview of the current Flavour Lattice Averaging Group results for the various nucleon isovector charges.}
%     \begin{tabular}{c|ccccc}
%         Method & $g_V$ & $g_A$ & $g_S$ & $g_T$ & $g_P$\\
%         \hline
%         FLAG 21 ($N_F=2+1$) & - & 1.248(23) & 1.13(14) & 0.965(61) & -\\
%         FLAG 21 ($N_F=2+1+1$) & - & 1.246(28) & 1.02(10) & 0.989(34) & - \\
%         CVC/PCAC & 1 & - & 1.02(10) & - & $-349(9)$
%     \end{tabular}
%     \label{tab:nucleon_charges}
% \end{table}

\subsection{Isospin symmetry breaking on $g_V$}
\label{sec:gV_isospin_breaking}
Traditionally, calculations of isospin symmetry breaking effects (due to quark mass differences and the electromagnetic interaction) have focused on the nuclear system where effects are expected to be sizeable (see Sec. \ref{sec:nuclear_structure}). For the individual nucleon, corrections to $g_V$ to due isospin symmetry breaking only show up at second order in the ISB interaction due to Behrens-Sirlin \cite{Behrends1956} and Ademolo-Gatto \cite{Ademollo1964} (BSAG) theorem, i.e. $g_V = 1 + \mathcal{O}(V_{C}^2)$ for isospin-breaking component of the Hamiltonian, $V_C$. Strong interaction isospin breaking effects arise from quark mass differences, so that corrections to $g_V$ were initially estimated to be of order $(m_u-m_d)^2/\Lambda_{QCD}^2 \sim 10^{-5}$, i.e. beyond the current experimental precision. Recent work, however, has cast some doubt upon this claim as some authors find corrections on the order several parts in $10^{-4}$, depending on their model of choice \cite{Crawford2022}. Such an effect would directly translate into shifts of the $V_{ud}$ value for all extractions on the order of its current uncertainty estimate, and be of significant importance. Its calculation is typically extremely difficult, however, as one usually looks for small perturbations in the Fermi matrix element due to the quark mass differences. Instead, the authors of Ref. \cite{Seng2023e} propose a method to calculate the ISB contribution directly, thereby reducing the required relative precision. They do so by appealing to a spectral decomposition of the nucleon excited state spectrum. The isospin ladder operator turning neutrons into protons and vice versa is isovector, so that one may perform a summation over all possible $T=1/2$ and $T=3/2$ states with the same quantum numbers and find for $g_V = 1 + \delta_C$
\begin{equation}
    \delta_C = \frac{(\Delta m_q)^2}{3}\left[\sum_a\frac{|\langle a;1/2 || \hat{O}^1 || N(s)\rangle|^2}{(m_N^0-E_{a,1/2})^2}-\sum_a \frac{|\langle a;3/2 || \hat{O}^1 || N(s)\rangle|^2}{(m_N^0-E_{a,3/2})^2}\right]
\end{equation}
where $\Delta m_q = m_u-m_d$, $\hat{O}^1$ is a rank-one tensor operator in isospin space, $N(s)$ is a nucleon state with spin $s$ with mass $m_N^0$ when isospin is conserved. Enhancements such as those found in Ref. \cite{Crawford2022} can then be understood as a coherent sum of individual excited states, thereby exceeding the simple power counting result. These may be calculated on the lattice directly, so that a large theory uncertainty on the non-perturbative calculation should result in a theory uncertainty below $10^{-4}$.

\subsection{Non-perturbative radiative corrections to $g_V$}
\label{sec:non_perturbative_gV}
All charges are renormalized at the nucleon and electron mass scale due to their interaction with other Standard Model fields. The dominant corrections arise from soft photon final state (i.e. Coulomb) interactions encoded primarily in the Fermi function, and logarithmic terms of order $\alpha \log M_Z/\mu$ of Eq. (\ref{eq:C_beta_r}). Despite this, the majority of the current theoretical uncertainty to RC of $g_V$ comes from the so-called $\gamma W$ box diagram, which arises from an interference with the axial weak current and is sensitive to non-perturbative excitations of the nucleon. As such, it has been the central object of research for an accurate determination of $V_{ud}$. The total energy-independent radiative correction may be written in the traditional description as \cite{Czarnecki2004}
\begin{equation}
    \Delta_R^V = \frac{\alpha}{2\pi}\left[3\ln \frac{M_Z}{m_N}+\ln \frac{M_Z}{M_W} + \tilde{a}_g\right] + \delta_\mathrm{HO}^\mathrm{QED} + 2 \square_{\gamma W}^{VA}.
\end{equation}
where $\tilde{a}_g = -0.083$ the $\mathcal{O}(\alpha_s)$ corrections to one-loop diagrams except for the axial $\gamma W$ box, and $\delta_\mathrm{HO}^\mathrm{QED} = 0.00109(10)$ contributions from leading-log higher-order QED corrections.

Without a clear means by which to calculate the non-perturbative parts of the diagram, the authors of Ref. \cite{Marciano2006} proposed an interpolation between the Born amplitude and perturbative QCD at high $Q^2$ with a 100\% relative uncertainty in 2006. In 2018, work by Seng, Gorchtein, Ramsey-Musolf and Patel \cite{Seng2018} showed that one can make progress using dispersion relations to connect problematic parts of the calculation to experimentally measured cross sections. As a first step, they showed how $\gamma W$ box diagram contribution to $\Delta_R^V$ may be written as
\begin{equation}
    \square_{\gamma W}^{VA} = \frac{e^2}{2Mg_V} \int \frac{d^4q}{(2\pi)^4} \frac{M_W^2}{M_W^2+Q^2}\frac{1}{Q^4}\frac{\nu^2+Q^2}{\nu}T_3(\nu, Q^2) \label{eq:box_gammaW_V}
    %\square_{\gamma W}^A &=& \frac{e^2}{Mg_A^\circ} \int \frac{d^4q}{(2\pi)^4} \frac{M_W^2}{M_W^2+Q^2}\frac{1}{Q^4}\left\{\frac{\nu^2-2Q^2}{3\nu}S_1-\frac{Q^2}{\nu}S_2\right\} \label{eq:box_gammaW_A}
\end{equation}
where $\nu = p\cdot q / M$, $Q^2= - q^2$ and $T_3(\nu, Q^2)$ is the spin-independent parity-violating structure function of the generalized Compton tensor, $T^{\mu\nu}_{\gamma W}$, i.e.
\begin{equation}
    T^{\mu\nu}_{\gamma W} \equiv \int d^4 x~ e^{iq\cdot x} \langle p | T[J^\mu_{em}(x)J_W^{\nu}(0)] | n \rangle \stackrel{\mathrm{asy}}{=} -\frac{i\epsilon^{\mu\nu\alpha \beta}q_\alpha p_\beta}{2p\cdot q}T_3(\nu, Q^2)% + \frac{i\epsilon^{\mu\nu\alpha\beta}q_\alpha}{p\cdot q}\left[S_\beta S_1 + \left(S_\beta-\frac{S\cdot q}{p\cdot q}\right)S_2\right]
\end{equation}
The initial work by Seng et al. \cite{Seng2018, Seng2019b} used dispersion relations to attempt a translation of the isoscalar\footnote{This refers to the isoscalar component of the electromagnetic interaction, shown by Sirlin \cite{Sirlin1967} to be the sole contributor.} $T_3^{(0)}$ component of the generalized Compton tensor into the analogous structure function of the \textit{full} hadronic tensor, $F_3^{(0)}$, where
\begin{equation}
    W^{\mu\nu}_{\gamma W} = \frac{1}{8\pi}\sum_X (2\pi)^4 \delta^4(p+q-p_X)\langle p | J^\mu_{em, 0} | X \rangle \langle X | J_W^\nu | n \rangle \stackrel{\mathrm{asy}}{=} \frac{i\epsilon^{\mu\nu\alpha \beta}p_\alpha q_\beta}{2p\cdot q}F_3^{(0)}
    \label{eq:W_munu_F3}
\end{equation}
for all intermediate states $X$. Their central result was to write the $\gamma W$ box diagram in terms of the first Nachtmann moment of $F_3^{(0)}$, noted $M_3^{(0)}(1, Q^2)$, 
\begin{eqnarray}
    \square_{\gamma W}^{VA} &=& \frac{3\alpha}{2\pi} \int_0^\infty \frac{dQ^2 M_W^2}{Q^2[M_W^2+Q^2]}M_3^{(0)}(1, Q^2) \quad \mathrm{with} \label{eq:box_Nachtmann} \\
    M_i(N, Q^2) &=& \frac{N+1}{N+2}\int_0^1\frac{dx \xi^N}{x^2}\left[2x-\frac{N\xi}{N+1}\right]F_i \label{eq:Nachtmann_def}
\end{eqnarray}
with $\xi = 2x/(1+\sqrt{1+4M^2x^2/Q^2})$ for $x=Q^2/(2M\nu)$. As $Q^2 \to \infty$ the one has $\xi\to x$ and the Nachtmann moment reduces to the regular Mellin moment. One benefit of the Nachtmann moment is the automatic inclusion of target mass corrections at low $Q^2$.

Experimental data was obtained from parity-violating neutrino-nucleus scattering data, from which one can obtain $F_3^{WW}$, i.e. both currents in Eq. (\ref{eq:W_munu_F3}) are $J_W^\mu$. The resonance structure of $F_3^{(0)}$ and $F_3^{WW}$ are different, however, such that the latter needed a deconstruction into its different components which could then individually be translated. The latter implies some model-dependence on their final result. The non-perturbative parts were described using Regge theory, and it was argued using vector meson dominance that both processes could be directly related. The resultant increase of the non-perturbative contribution compared to the 2006 work resulted in an 2.3 $\sigma$ increase (decrease) in $\Delta_R^V$ ($V_{ud}$).

Another approach has been to rely on QCD sum rules and parametrizations of the non-perturbative parts using, e.g., holographic QCD. This was performed by the reanalysis of their 2006 result in a paper by Czarnecki, Marciano and Sirlin, confirming the substantial increase found by Seng et al \cite{Czarnecki2020}. A remaining discrepancy between their result and that using dispersion relations was found to be due to target-mass corrections in a work by Hayen \cite{Hayen2021}. Finally, a similar analysis by Shiells et al. \cite{Shiells2021} confirmed the initial calculations of Seng et al. and the field has reached a reasonable degree of consensus. Several compilations have recently been published to put these results of equal footing and arrive as an average that should serve at the community standard. We take here the summary by Cirigliano et al. \cite{Cirigliano2023} and write down
\begin{equation}
    \Delta_{R}^V = 0.02467(27)
\end{equation}

\begin{figure}[ht]
    \centering
    \includegraphics[width=\textwidth]{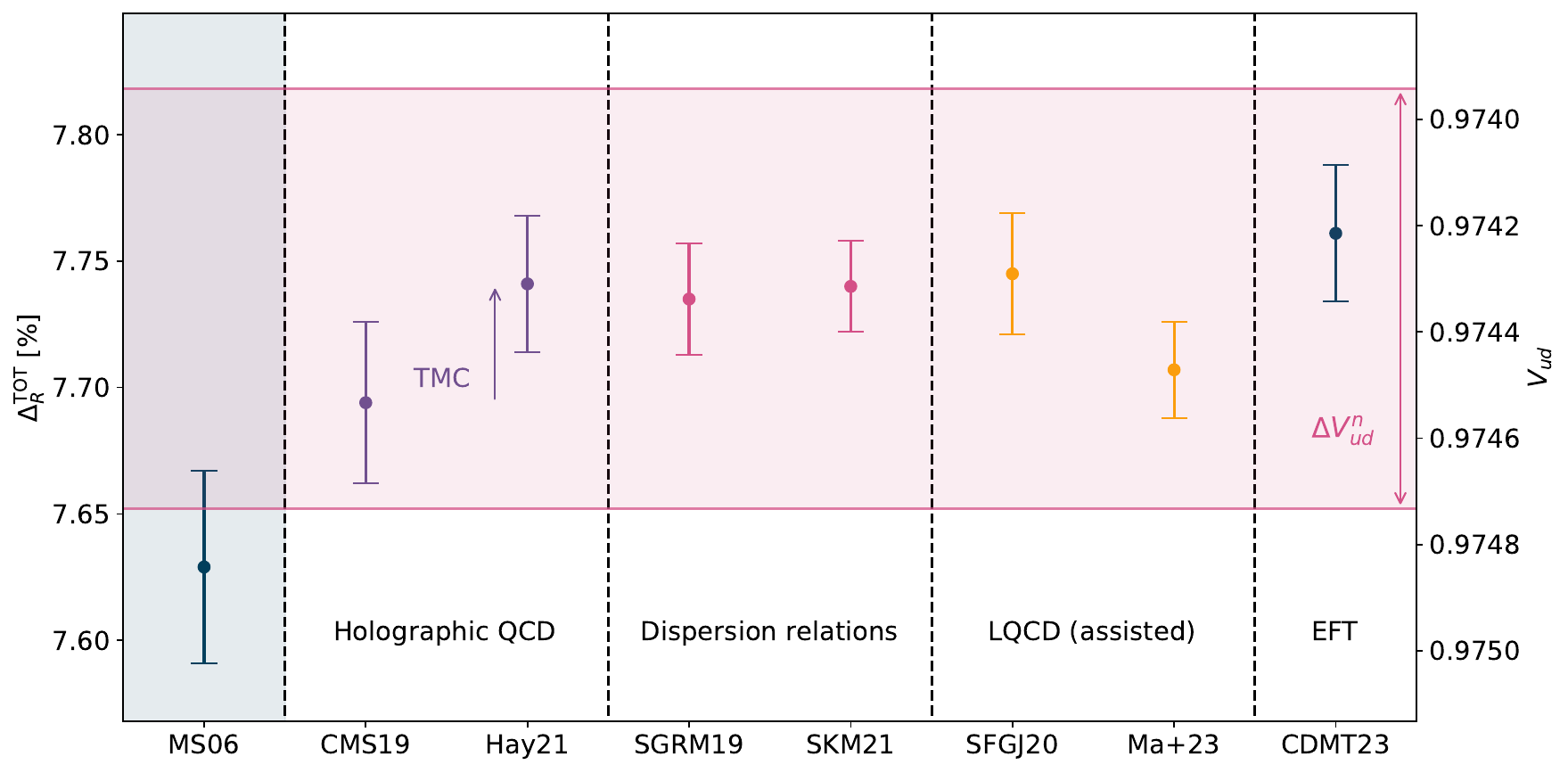}
    \caption{Summary of recent calculations of radiative corrections to the neutron decay rate. Following the original result by Marciano and Sirlin (MS06) \cite{Marciano2006}, in which non-perturbative effects were incorrectly estimated, progress has been made using different techniques: Holographic QCD with a reanalysis by Czarnecki \textit{et al.} (CMS19) \cite{Czarnecki2019} and Hayen (Hay21) \cite{Hayen2021} including target mass corrections (TMC); Dispersion relations using experimental neutrino-nucleon scattering by Seng \textit{et al.} (SGRM19) \cite{Seng2019} and Shiells \textit{et al.} (SKM21) \cite{Shiells2021}; using $\chi$EFT to use LQCD pion results (SFGJ20) \cite{Seng2020} and a LQCD nucleon calculation by Ma \textit{et al.} (Ma+23) \cite{Ma2023}; and an EFT approach in in which all final state corrections were discussed by Cirigliano et al. (CDMT23) \cite{Cirigliano2023}.}
    \label{fig:summary_RC}
\end{figure}

In recent years, lattice QCD results have progressed through successively more difficult calculations to arrive at a first lattice determination of the box diagram for the pion \cite{Feng2020}, which were successively applied to the nucleon using $\chi$EFT \cite{Seng2020}. The first calculations using baryons on the lattice were performed only recently, and find results consistent if somewhat lower than the dispersive analyses \cite{Ma2023}. In this latest result, $\Delta_R^V = 0.02439(19)$, the remaining uncertainty arises predominantly from the continuum extrapolation for $Q^2 < 2$ GeV$^2$ and higher-loop truncation effects for $Q^2 > 2$ GeV$^2$. The total uncertainty is now smaller than the experimental one for superallowed decays, even though the latter's uncertainty estimate is currently dominated by nuclear structure effects (see Sec. \ref{sec:nuclear_structure}). Figure \ref{fig:summary_RC} shows a summary of the current total radiative corrections for the neutron, including that following a reanalysis of the Fermi function and higher-order radiative corrections discussed in Sec. \ref{sec:higher_order_QED}, with $\Delta_R^\mathrm{TOT}$ defined as
\begin{equation}
    |V_{ud}|^2\tau_n(1+3\lambda^2)(1+\Delta_R^\mathrm{TOT}) = 5283.321(5)~ s,
\end{equation}
as in Ref. \cite{Cirigliano2023a} and includes the effect of the Fermi function.

\subsection{Non-perturbative radiative corrections to $g_A$}
\label{sec:gA_radiative_corrections}
Like the vector case, the axial charge is renormalized by strong and electroweak interactions. Changes due to the former are not anticipated to be small, however, as the non-conservation of the axial current does not require the cancellation of diagrams as in the vector case. Other than the dominant leading-log terms of Eq. (\ref{eq:C_beta_r}), work on $\Delta_R^A$ was limited as its precise value was not necessary for extractions of $V_{ud}$. In fact, one defined an effective $g_A$ through the neutron lifetime by writing \cite{Czarnecki2004}
\begin{equation}
    |V_{ud}|^2(1+3\tilde{g}_A^2)\tau_n = 4908(4)~s
\end{equation}
where
\begin{equation}
    \tilde{g}_A = g_A^\mathrm{QCD}\left[1+\frac{1}{2}(\Delta_R^A-\Delta_R^V) + \delta_{BSM}\right]
    \label{eq:gA_eff}
\end{equation}
where $\delta_{BSM}$ are potential Beyond Standard Model signatures. Using the LEFT prescription of Eq. (\ref{eq:linearized_LEFT}), we have $\delta_{BSM} = -2\mathrm{Re}~\epsilon_R^\prime$, meaning that sufficiently precise LQCD calculations of $g_A$ are able to probe right-handed currents directly. On the other hand, the definition of $g_A$ as in Eq. (\ref{eq:gA_eff}) is useful only if the same $\tilde{g}_A$ can be extracted from experiment, and the difference $\Delta_R^A-\Delta_R^V$ can be calculated.

The study of $\Delta_R^A$ has only recently been undertaken, with initial works by Hayen using holographic QCD methods \cite{Hayen2021} and Gorchtein \& Seng using dispersion relations \cite{Gorchtein2021}. Since then, a number of different works have appeared and while the contributions from the $\gamma W$ box are reasonably well understood, unexpectedly large changes were found due to enhanced isospin breaking effects within the $\mathcal{O}(\alpha)$ calculation. This remains an open problem, and likely needs dedicated lattice QCD calculations, discussed below.

\subsubsection{Contributions from $\gamma W$ box}
Like the vector case above, the contribution to $\Delta_R^A$ from the $\gamma W$ box may be written as follows
\begin{equation}
\square_{\gamma W}^{VV} = \frac{e^2}{Mg_A} \int \frac{d^4q}{(2\pi)^4} \frac{M_W^2}{M_W^2+Q^2}\frac{1}{Q^4}\left\{\frac{\nu^2-2Q^2}{3\nu}S_1-\frac{Q^2}{\nu}S_2\right\} 
\label{eq:box_gammaW_A}
\end{equation}
where now the vector component of the weak interaction can give rise to a contribution to the axial current. Because of the spin-dependence of the axial current, the result depends on the spin-dependent structure functions,
    \begin{equation}
    T^{\mu\nu}_{\gamma W} \equiv \int d^4 x~ e^{iq\cdot x} \langle p | T[J^\mu_{em}(x)J_W^{\nu}(0)] | n \rangle \stackrel{\mathrm{spin}}{=}  \frac{i\epsilon^{\mu\nu\alpha\beta}q_\alpha}{p\cdot q}\left[S_\beta S_1 + \left(S_\beta-\frac{S\cdot q}{p\cdot q}\right)S_2\right]
\end{equation}
where $S_\beta$ is the nucleon spin normalized using $S^2=-m_N^2$. The work by Hayen \cite{Hayen2021} evaluated the non-perturbative elements of the integral using a combination of holographic QCD and the experimental measurements of polarized Bjorken sum rule, and adding target mass corrections \textit{a posteriori} together with higher-twist effects, finding $\Delta_R^A-\Delta_R^V = 0.60(5) \times 10^{-3}$. Later, Gorchtein and Seng \cite{Gorchtein2021}, used a dispersion relation to transform the integral of Eq. (\ref{eq:box_gammaW_A}) into Nachtmann moments of polarized nucleon structure functions, $g_{1, 2}$. The $g_1$ contribution, for example, depends on
\begin{equation}
    \overline{\Gamma}_1^{p-n}(Q^2) = \int_0^{x_\pi} dx_B \frac{4(5+4r)}{9(1+r^2)}(g_1^p(x_B, Q^2)-g_1^n(x_B, Q^2))
    \label{eq:nachtmann_g1}
\end{equation}
with $x_B = Q^2/(2p\cdot q)$ the Bjorken-$x$ and $r=\sqrt{1+4m_N^2x_B^2/Q^2}$. For $Q^2 \to \infty$ the Nachtmann moments reduce to the Mellin moments measured in deep inelastic scattering (DIS),
\begin{equation}
    \Gamma^N_i(Q^2) = \int_0^{x_\pi} dx_B x_B^{i-1}g_1^N(x_B, Q^2)
\end{equation}
so that Eq. (\ref{eq:nachtmann_g1}) may be obtained through a linear combination of Mellin moments constrained by data \cite{Gorchtein2021}. The benefit of this approach is the automatic inclusion of target mass corrections and the direct connection to data thanks to isospin symmetry. Additionally, they provided an estimate for the contribution of $g_2$ and find a result of $\Delta_R^A-\Delta_R^V = 0.13(13) \times 10^{-3}$. This result was recently confirmed using an explicit lattice QCD calculation which reported $\Delta_R^A-\Delta_R^V = 0.07(11) \cdot 10^{-3}$ \cite{Ma2023}.

\subsubsection{Enhanced isospin breaking corrections}
Contrary to the vector coupling, the axial vector current is not conserved and receives additional corrections due to 
%a different diagram through
% \begin{equation}
%     \mathcal{M}^\gamma_v \to \frac{g^2e^2}{4(2\pi)^4}V_{ud}\frac{L^\mu}{q^2-M_W^2}\lim_{\bar{q}\to q} \frac{\partial}{\partial \bar{q}} \mathcal{D}_\gamma
% \end{equation}
%where $\mathcal{D}_\gamma$ depends on the 
the non-conservation of the  axial weak current in a three-point function
\begin{equation}
    \mathcal{D}_\gamma = \int \frac{d^4 k}{k^2}\int d^4 y e^{i\bar{q}y}\int d^4x e^{ikx} \langle p_f | T\{\partial_\mu J^\mu_W (y) J_{em}^\lambda(x) J_{\lambda}^{em}(0)\}| p_i \rangle
    \label{eq:D_three_point}
\end{equation}
In Ref. \cite{Hayen2021} this contribution was evaluated in the asymptotic limit and the Born contribution using the partially conserved vector current and the Goldberger-Treiman relationship to show that both contributions vanished in the isospin limit.

A new analysis using chiral perturbation theory \cite{Cirigliano2022a}, however, showed that isospin breaking corrections are strongly enhanced. In particular, the electromagnetic mass splitting of the pion was shown to give large, percent-level corrections to the renormalized value of $g_A$ obtained in experiments. This effect was found through a matching procedure between different types of effective field theories. In particular, by matching heavy baryon chiral perturbation theory (HB$\chi$PT) and pion-less EFT ($\slashpi$ EFT) one could isolate effects these particular effects in a consistent expansion using Eq. (\ref{eq:g_V_A_CHIPT_expansion}).
% and write down
% \begin{equation}
%     g_{V/A} = g_{V/A}^{(0)} \left[1+\sum_{n=2}^\infty \Delta_{V/A,\chi}^{(n)} + \frac{\alpha}{2\pi}\sum_{n=0}^\infty \Delta_{V/A, em}^{(n)} + \left(\frac{m_u-m_d}{\Lambda_\chi}\right)^{n_{V/A}}\sum_{n=0}^\infty\Delta_{V/A, \delta m}^{(n)}\right]
% \end{equation}
% where correction terms correspond to chiral corrections ($\chi$), electromagnetic and strong isospin symmetry breaking. 
Surprisingly, both at leading and next-to leading order one obtained percent-level corrections proportional to the pion-mass splitting encoded in $Z_\pi = 0.81$
\begin{eqnarray}
    \Delta_{A, em}^{(0)} &=& Z_{\pi} \left[\frac{1+3g_A^{(0)2}}{2}\left(\log \frac{\mu^2}{m_\pi^2}-g_A^{(0)2}\right)\right] + \hat{C}_A(\mu) \\
    \Delta_{A, em}^{(1)} &=& 4 Z_{\pi}\pi m_\pi \left[c_4-c_3+\frac{3}{8m_N}+\frac{9}{16m_N}g_A^{(0)2} \right]
\end{eqnarray}
where $\mu$ is a renormalization scale, $c_{3, 4}$ are low energy constants which can be obtained from pion-nucleon scattering and $g_{V, A} = C^r_\beta[1+(\alpha/2\pi)\hat{C}_{V, A}]$. The influence of the leading order was only estimated as the $\mu$ dependence in $\hat{C}_A$ is not yet known, and as a first approximation one set $\hat{C}_A-\hat{C}_V = 0$ and varied $\mu$ between 0.5 and 1 GeV. As this shows up in the leading order result, this translates into a sizeable uncertainty in the electromagnetic renormalization of $\lambda \equiv g_A/g_V$, which is renormalization-scale independent
\begin{equation}
    \lambda = \lambda_{QCD} \left[1+ 0.020(6)_{3pt} + 0.0001(1)_{\gamma W} - 2\mathrm{Re}~\epsilon_R^\prime\right]
    \label{eq:lambda_numerical_final}
\end{equation}
This result has no direct effect on the neutron or mirror $V_{ud}$ extraction as experiment determines a similarly renormalized $\lambda$ value. It does, however, have an immense effect on the comparison between lattice QCD and experimental determinations of $g_A$ to search for exotic right-handed currents.

%The isospin breaking corrections also serve to shift the value of weak magnetism (see App. \ref{app:recoil_order}) in the nucleon system, which in turn gives rise to changes in the $\beta$-$\nu$ angular correlation and $\beta$-asymmetry \cite{Cirigliano2022a}. For the neutron these shifts are $\mathcal{O}(10^{-4})$ and might become important for upcoming experiments.

\subsubsection{Lattice QCD + QED$_M$ calculations}
In order to resolve the uncertainty in Eq. (\ref{eq:lambda_numerical_final}) one may calculate its effect directly on the lattice. The latter is technically extremely difficult if one were to calculate the three-point function (Eq. (\ref{eq:D_three_point})) as on the lattice this becomes a five-point function through the nucleon source and sink operators as discussed in Sec. \ref{sec:bare_charges_QCD}. Instead, one may attempt to calculate $\langle p_f | J^\mu_W | p_i \rangle$ with photons on the lattice. While this is typically difficult due to the infinite range of the photon compared to the finite volume of numerical calculations, an approach known as QED$_M$ is able to overcome these difficulties by introducing a finite photon mass as an IR regulator \cite{Endres2016}. Such a theory is renormalizable, local, and can be implemented in existing codes with minimal modification \cite{Clark2022}. In order to perform the calculation for, e.g., $\Delta_R^A$, one must know the individual corrections on the lattice to pick apart from the total $n\to p+e^-+\bar{\nu}$ amplitude. This requires a number of non-trivial calculations, such as the finite volume corrections, the modification of the electron propagator in the QED background, and the non-perturbative renormalization of the calculation \cite{Walker-LoudPC}. In order to reach the required precision for a determination of Eq. (\ref{eq:lambda_numerical_final}), substantial computing power is likely required.

\subsection{Higher-order QED corrections}
\label{sec:higher_order_QED}
Up to now, the total integrated decay rate for neutrons and nuclei, as in Eq. (\ref{eq:ft_mixed}), consist of pieces that have typically been calculated using different theoretical frameworks. In the traditional method, the decay rate is written as follows, (see Eq. (\ref{eq:spectrum_shape_general}))
\begin{equation}
    \Gamma \propto \int_1^{W_0} dW p W (W_0-W)^2 F(Z, W) C_{V/A}(Z, W, W_0) K(Z, W, W_0)[1+\delta_{RC}].
\end{equation}
The calculation of the Fermi function is typically performed by considering the long wavelength limit of virtual photon exchange and instead replacing the electron wave function from a plane wave to the solution of the Dirac equation in a static Coulomb potential. This has the advantage of being analytically tractable and resumming many of the dominant $(\alpha Z)^n$ corrections, with other radiative corrections calculated explicitly using diagrammatic methods. Additional terms arising from other photon-exchange diagrams are contained in $\delta_{RC}$ to make the product $F(1+\delta_{RC})$ agree with the perturbative results up to the order of the calculation. Recently, however, this methodology is under scrutiny, as approaches using effective field theory find different results and stress the need for proper matching between scales in consistent renormalization schemes. Specifically, the differences appear in the large-log summation and the treatment of the Fermi function, which we discuss separately.

\subsubsection{RGE and large-logarithm resumming}
The vector coupling constant was recently discussed within an EFT framework going from the Standard Model to Low-Energy Effective Field Theory below the weak scale (LEFT) to heavy baryon chiral perturbation theory at the nucleon scale including pions and finally finally pion-less EFT at the electron mass \cite{Cirigliano2023a}. We largely base on discussion on their results and \cite{DekensPC, MereghettiPC}, and focus here on the RGE evolution to resum the large logarithms (see App. \ref{app:renormalization}). Their schematic result for $g_V$ is the following
\begin{equation}
    g_V(m_e) = U(m_e, \mu_\chi) \left[1+\square_\mathrm{Had}^V+\frac{\bar{\alpha}(\mu_\chi)}{\pi}\kappa\right]U(\mu_\chi, \mu_W)C_\beta^r(\mu_W)
\end{equation}
where both $U$'s originate from solving the renormalization group equations to next-to-leading order to connect relevant scales, and $\square_{Had}^V$ is proportional to the $\gamma W$ box contributions discussed above. Differences with the literature appear at second order in the RGE's of both the LEFT to HBChPT resumming and the HBChPT to the electron mass scale. Specfically, in the RGE for the Wilson coefficient $C_\beta^r$ Ref. \cite{Cirigliano2023a} finds a difference in $\gamma_1$, the $\mathcal{O}\{(\alpha/\pi)^2\}$ running coefficient, which is 5.5 times smaller than that of Ref. \cite{Czarnecki2004}. The origin of the difference is not identified, and the change in resummed $\alpha^2 \ln(M_W/m_c)$ terms reduces $\Delta_R$ by 0.011\%. 

Additional important differences in the RGE resumming, however, originate at $\mathcal{O}(\alpha^2\ln (m_N/m_E))$, i.e. running $g_V$ from $\mu=\mu_\chi\sim m_N$ to $\mu=m_e$. Differences with respect to the literature appear in the two-loop anomalous dimension. Traditionally, these have been calculated using relativistic nucleons by Jaus \cite{Jaus1970}, and Jaus and Rasche \cite{Jaus1972} and remained untouched since the late 1980's \cite{Jaus1987, Sirlin1986}. The calculations were performed to order $\mathcal{O}(\alpha^2 Z)$ including finite parts, but approximated one of the photons as a Coulomb interaction. The work of Ref. \cite{Cirigliano2023a} calculated only the contributions to the two-loop RGE (i.e. the divergent parts) enhanced by factors of $\pi^2$, but were unable to obtain agreement with traditional calculations \cite{DekensPC}. An identical observation was made in Ref. \cite{Hill2023}, discussed in more detail below. This difference resulted in a positive shift of $\Delta_R^V$ of $0.010\%$.

For finite $Z$, the results by Hill and Plestid \cite{Hill2023, Hill2023a} provide new results for the $Z$-dependent RGE up to three-loop order. If confirmed, these substantially improve upon the heuristic estimates of order $\mathcal{O}(\alpha^3 Z^2)$ \cite{Sirlin1987b} and the simple approximations at higher orders \cite{Wilkinson1997}.

\subsubsection{Fermi function} As mentioned above, the Fermi function is typically calculated using a solution of the Dirac equation in a central Coulomb potential up to all orders in $\alpha Z$, and additional terms from a full perturbation theory calculation put into $\delta_{RC}$ to find \cite{Hayen2018}
\begin{equation}
    F(Z, W, R) = 2(1+\gamma)(2pR)^{2(\gamma-1)}e^{\pi y}\frac{|\Gamma(\gamma+iy)|^2}{|\Gamma(1+2\gamma)|^2}
\end{equation}
where $R=\sqrt{5/3}\langle r^2 \rangle_\mathrm{exp}^{1/2}$ represents the nuclear radius and $y=\alpha Z/\beta$ for $\beta = p/E$ the electron velocity. To first order in $\alpha Z$ one finds $F \sim 1 + y$.

Besides the EFT construction discussed above by Ref. \cite{Cirigliano2023a}, recent efforts to consistently calculate the enhanced Coulomb corrections were reported in Refs. \cite{Hill2023, Hill2023a}. The argument used in the latter works is the appearance of an unphysical UV regulator in the form of the nuclear radius, $R\sim 1/\mu_\chi$, in $F(Z, W)$. This is introduced in the traditional calculation as a cut-off for the divergent solution of the Dirac equation of a point charge at the origin, the argument being that the electron wave function changes little inside the nuclear volume. One then introduces `correction terms' denoted as $L_0$ for a finite-size charge distribution derived either analytically or numerically and additionally includes atomic screening corrections ($S$), shape-dependent terms ($U$), $\ldots$ \cite{Behrens1982, Wilkinson1993b, Hayen2018}, by constructing more elaborate potentials.

In keeping with a consistent renormalization scheme, however, Refs. \cite{Cirigliano2023a, Hill2023, Hill2023a} construct an EFT approach to capture the resummed ($\pi \alpha/\beta$)$^n$ effects. Matching $\slashpi$EFT onto existing non-relativistic QED calculations, Ref. \cite{Cirigliano2023a} argues for the use of the non-relativistic Fermi function at $\mu=m_e$,
\begin{equation}
    F_{NR}(\beta) = \frac{2\pi \alpha}{\beta} \frac{1}{1-e^{-\frac{2\pi \alpha}{\beta}}}
    \label{eq:F_NR}
\end{equation}
up to finite pieces of $\mathcal{O}(\alpha^2)$ (see above). The work by Refs. \cite{Hill2023, Hill2023a}, on the other hand, uses all-order factorization theorems to reconstruct an object very close to the relativistic Fermi function with an explicit renormalization scale dependence, $\mu_H$, as
\begin{equation}
    F^\prime(Z, E)(\mu_H) = F(Z, E, r_H) \frac{4\gamma}{(1+\gamma)^2}
\end{equation}
with $r_H^{-1} = \mu_He^{\gamma_E}$ a radius. We may connect this to the result of Eq. (\ref{eq:F_NR}) by writing an expansion of the relativistic Fermi function as \cite{Cirigliano2023a}
\begin{equation}
    F_0(\beta) = F_{NR}(\beta)[1-\alpha^2(\gamma_E-3+\ln(2E_e R \beta)) + \mathcal{O}(\alpha^4)].
\end{equation}
Setting $\mu_H = m_e$ corresponds to $R = (m_ee^{\gamma_E})^{-1}$ and
\begin{equation}
    F^\prime(m_e) = F_{NR}(\beta)[1-\alpha^2(-3+\ln(2W\beta))+\mathcal{O}(\alpha^4)]
    \label{eq:F_prime_to_NR}
\end{equation}
It is now interesting to note that the finite (i.e. not logarithmically enhanced) $\mathcal{O}(\alpha^2Z)$ terms as calculated numerically by Jaus and Rasche \cite{Jaus1970, Jaus1972} were approximated analytically by Sirlin and Zucchini \cite{Sirlin1986, Czarnecki2004} as
\begin{eqnarray}
    \delta_{\alpha^2Z}^\mathrm{fin} &\stackrel{ERA}{=}& \alpha^2 Z\left[\frac{5}{3}\ln(2W)-\frac{43}{18}\right] \\
    &\stackrel{NRA}{=}& \alpha^2 Z\left[\frac{2}{3}\ln(2W) - \frac{35}{9}-\frac{\pi^2}{6} + 6 \ln 2\right]
\end{eqnarray}
in the extreme relativistic approximation (ERA, $\beta\to 1$) and non-relativistic approximation (NRA), are very similar to that obtained with Eq. (\ref{eq:F_prime_to_NR}). While it was noted in Ref. \cite{Cirigliano2023a} that numerically these largely cancelled the effect of using the non-relativistic Fermi function in their calculation, we may provide a deeper understanding using the results of Refs. \cite{Hill2023, Hill2023a}.

\subsubsection{Summary}
We may therefore attempt to summarize the current status as follows: differences at the two-loop anomalous dimensions remain for both LEFT to HBChPT and HBChPT to $m_e$ RGE's, but introduce $\mathcal{O}(0.01\%)$ corrections to $\Delta_R$ which almost completely cancel. Up to finite pieces of $\mathcal{O}(\alpha^2)$, the $\slashpi$EFT calculations of Ref. \cite{Cirigliano2023a} agree with those of Refs. \cite{Hill2023, Hill2023a}. Using a consistent renormalization scheme, the EFT results of the latter largely reproduce the finite pieces of the traditional calculations using a relativistic Fermi function for a physical radius $R$ and $\mathcal{O}(\alpha^2 Z)$ corrections, questioning the discrepancy found in Ref. \cite{Cirigliano2023a}. The exact correspondence for the neutron as well as for finite $Z$ needs to be further investigated.

\section{NUCLEAR STRUCTURE CORRECTIONS}
\label{sec:nuclear_structure}

For the past decades, both the most precise determination of $V_{ud}$ and the strongest constraint on exotic scalar currents came from superallowed $0^+$ to $0^+$ decays, through a combination of an impressive experimental data set and significant efforts to address a myriad of nuclear structure corrections. Recently, however, changes to nuclear-structure corrections to the $\gamma W$ box ($\delta_{NS}$) have significantly altered the numerical input to Eq. (\ref{eq:ft_mixed}) and dominate the $V_{ud}$ uncertainty budget \cite{Cirigliano2023}
\begin{equation}
    V_{ud}^{0^+\to 0^+} = 0.97367(11)_\mathrm{exp}(13)_{\Delta_R^V}(27)_\mathrm{NS}[32]_\mathrm{total}
\end{equation}
Besides this, renewed attention is going towards an evaluation of isospin breaking corrections and the calculation of the phase space factors. The former has a long history, with a wide variety of models of different levels of sophistication. In the works by Towner and Hardy \cite{Hardy2009, Hardy2020}, the equality of all $\mathcal{F}t$ values with a good $\chi^2/\nu$ was used as a prerequisite.
%\footnote{This was done since not all models \cite{Miller2008, Miller2009, Satula2011, Ormand1989, Ormand1995, Auerbach2009, Liang2009, Xayavong2018} have calculated uncertainties, in an attempt to put them on equal footing. This may disadvantage models, however, for which the calculated uncertainty is larger than the scatter in the calculated data.}
Recently, a new method was proposed to use electroweak charge radii to constrain isospin breaking effects \cite{Seng2023a, Seng2023b} which motivated a new measurement of the $^{26m}$Al charge radius \cite{Plattner2023}. Simultaneously, data-driven methods are used to reevaluate the calculated phase-space calculations and their uncertainty\footnote{We note that the use of weak charge radii rather than proton charge radii in the calculation of the phase space factors is well-known, and discussed for example as isovector corrections in Refs. \cite{Wilkinson1993b, Hayen2018}.} \cite{Seng2023c, Seng2023d}.

\subsection{Nuclear structure effects in the $\gamma W$ box}
\label{sec:delta_NS}
Towner and Hardy introduced the nuclear-structure radiative correction as the difference between the free-nucleon value and that modified by the nuclear medium. The approach was divided into two: $(i)$ corrections where the same nucleon was involved in both the weak contact interaction and the photon exchange with the final lepton, denoted $\delta_{NS}^A$; and $(ii)$ with photon and $W$ boson contact interactions on different nucleons, $\delta_{NS}^B$. In their original approach, the former was approximated by quenching the Born response in the same way that one phenomenologically quenched the value of $g_A$ for Gamow-Teller transitions between low-lying states, while the latter was estimated using free nucleons propagating inside a spectator nucleus with non-relativistic shell model codes.

The dispersion approach used first in the nucleon case allowed for two significant improvements in its description: ($i$) reasoning from the structure function picture, the quasi-elastic response was found to have the largest contribution to $\delta_{NS}$ rather than phenomenological quenching \cite{Seng2019}; and ($ii$) the low scales of nuclear excitations relative to nucleon excitations meant that energy-dependent terms contributed non-negligibly for nuclear transitions \cite{Gorchtein2018}. 

Recently, a more complete work covering additional corrections in a dispersion relation framework was published \cite{Seng2023}. We may briefly summarize their work by highlighting one of their central results,
\begin{eqnarray}
    \mathcal{R}e\square^\mathrm{nucl}_{\gamma W} - \square_{\gamma W}^{n} &\approx& \frac{3\alpha}{2\pi}\int_0^\infty \frac{dQ^2}{Q^2}\left\{\left[M_{3,-}^\mathrm{nucl}(1, Q^2)-M_{3,-}^n(1, Q^2)\right]\right. \nonumber \\
    &+& \left. \frac{8E_e M}{9Q^2}\left[M_{3,+}^\mathrm{nucl}(2, Q^2)-\eta M_{1,+}^{\prime \mathrm{nucl}}(2, Q^2)-\frac{3}{2}\eta M_{2,-}^\mathrm{nucl}(1, Q^2)\right]\right\}
    \label{eq:box_differences_nucleon_nuclear}
\end{eqnarray}
where the additional sign subscript denotes the even or odd component under $\nu\to -\nu$ and $\eta = \pm 1$ for $\beta^\pm$ decays. Keeping in mind the results for the single nucleon in terms of Nachtmann moments (Eqs. \ref{eq:box_Nachtmann} \& \ref{eq:Nachtmann_def}), the first line trivially shows the difference in nuclear and nucleon Nachtmann moments of $F_3$. The second line is more interesting, however, as as additional $\beta$ \textit{energy-dependent} terms appears with sensitivity only to the \textit{nuclear} Nachtmann moments. This is to be understood intuitively as the breakdown of the separation of scales in the original approach by Sirlin \cite{Sirlin1967}, with corrections separated into inner and outer components, as now excitation energies inside the nuclear blob can be comparable to the outgoing energies \cite{Gorchtein2018} and terms of $\mathcal{O}(\alpha/\pi (E_e/\Lambda_\mathrm{nucl}))$ cannot be ignored.

It would be highly desirable to perform an experimental measurement of the energy dependence in parallel with theoretical estimates, but the smallness of the slope (on the order of a few parts in $10^4$ per MeV) makes this extremely challenging.

\subsubsection{Structure function differences: quasi-elastic knockout and shadowing}
The energy-independent term in Eq. (\ref{eq:box_differences_nucleon_nuclear}) depends on the difference in the structure functions (through the Nachtmann moment) of a nucleus versus that of a single nucleon. For relatively low momenta and energies, the nucleus has a much richer structure than that of the nucleon through the appearance of individual levels at the MeV-scale and collective excitations such as the Giant Dipole Resonance at tens of MeV. While these can be probed using traditional nuclear physics experiments, their behaviour inside the $\gamma W$ box cannot be easily translated and instead a dedicated nuclear \textit{ab initio} study must be performed. This is currently available only for the lowest mass nuclei such as $^{10}$C, where one attempts to calculate the full nuclear Green's function up to medium $Q^2$ \cite{Navratil2023}.

At medium $Q^2$ the nuclear response function shows a strong quasi-elastic component, broadened by the Fermi function of the nucleons inside the nucleus. Contrary to the work by Towner and Hardy, it was shown that the quasi-elastic response is the dominant contributor at low to medium $Q^2$ \cite{Seng2019}. The latter was calculated using a simple free fermion gas with Pauli blocking, and using average removal energies. Because of the simplified approach, the current estimate carried a large relative uncertainty but nonetheless changed $\delta_{NS}$ by about 3 standard deviations. The resultant $\mathcal{F}t$ values changed by about 1.5 s, while the uncertainty on the quasi-elastic contribution was taken to be fully correlated over all isotopes.

For the nucleon, the majority of the $\gamma W$ box contribution comes from the Deep Inelastic Scattering regime. In previous analyses, it was assumed that the nuclear and nucleon structure functions would coincide in this regime and therefore do not contribute do $\delta_{NS}$. It is well-known, however, that the DIS nuclear structure functions typically show shadowing and EMC effects \cite{Geesaman1995, Kopeliovich2013}. Experimentally, this has been studied predominantly for the parity-conserving structure function $F_2$ \cite{Aubert1983, AmeodoM1995, Seely2009} rather than the parity-violating function $F_3$ entering $\Delta_R^V$. Despite a long history, the precise origin of the EMC effect is still not understood even though a number of different models are able to reproduce (parts of) the experimental data. Initial studies on $F_3$ showed strong effects on the structure function but found large cancellations in the Gross-Llewellyn-Smith sum rule \cite{Kulagin1998}. To leading twist the latter informs the DIS contribution to $\Delta_R^V$, but it is not clear whether the same conclusion is reached in other models. Motivated by the physics program of DUNE, however, studies are gaining renewed attention \cite{Athar2021, Kulagin2007, Kulagin2010}. 

% \subsubsection{Energy-dependent terms}
% Of particular interest is the fact that the energy-dependent corrections in the second line appear only for nuclear systems, whereas the leading energy-independent terms depend on the difference of nuclear and nucleon Nachtmann moments $M_3$.

% So far, it has only been estimated using a free fermion gas, in which case the effect is dominated by $M_{3,+}^\mathrm{nucl}$ with the other two terms contributing only at the $10^{-5}$ level. This results in a transition-dependent correction in the superallowed set in a way that interferes with the scalar current extraction. It would be highly desirable to perform an experimental measurement of the energy dependence in parallel with theoretical estimates, but the smallness of the slope (on the order of a few parts in $10^4$ per MeV) makes this extremely challenging.

\subsection{Mirror transitions and nuclear structure effects}
\label{sec:mirror_delta_NS}
Despite the potential for mirror decays in extracting competitive determinations of $V_{ud}$, the evaluation of nuclear structure and isospin breaking corrections to the Gamow-Teller to Fermi mixing ratio has not been studied in detail. Specifically, the traditional analysis for extracting $V_{ud}$ with Eq. (\ref{eq:ft_mixed}) makes the following approximation for $\rho$,
\begin{equation}
    \rho \equiv \frac{g_A |M_{GT}|(1+(\delta_{NS}^A-\delta_C^A)/2)}{g_V |M_F|(1+(\delta_{NS}^V-\delta_C^V)/2)} \approx \frac{g_A |M_{GT}|}{g_V |M_F|}
    \label{eq:rho_def_approx}
\end{equation}
where $\rho$ is assumed to be the same as that obtained from measurements of angular correlation coefficients (see Sec. \ref{sec:available_experimental_observables}). As the Gamow-Teller matrix element can not be determined from first principles to a sufficient precision, this approximation is typically not given much attention. While the current mirror $V_{ud}$ determination (after removing double counting effects, see below) is consistent with that obtained from the neutron and the superallowed data set, it is by no means guaranteed that this should be the case.

\subsubsection{Structure function differences and enhanced isospin breaking}
Following the discussion on $\Delta_R^A$ in Sec. \ref{sec:gA_radiative_corrections} and $\delta_{NS}^V$ above, the approximation in Eq. (\ref{eq:rho_def_approx}) is unlikely to be correct. In particular, while the quasi-elastic contributions may be expected to be similar for $\delta_{NS}^{V, A}$, effects from discrete states are likely to differ. Similarly, nuclear (anti)shadowing effects play an important role in the DIS regime which dominates the $\gamma W$ contribution. For the axial current, its effect depends on the (anti)shadowing of the spin-dependent structure functions, $g_{1, 2}$ \cite{Kulagin1995}. While these have received less attention than their spin-independent partners, initial studies \cite{Thomas1995, Kaptari1995} have shown a stronger EMC effect for spin-polarized structure functions \cite{Cloet2005, Smith2005, Cloet2012a, Cloet2006}. The effects on the polarized Bjorken sum rule have not been explicitly researched, however, and require further study. 

As was the case for the neutron (Sec. \ref{sec:gA_radiative_corrections}), isospin breaking corrections due to the weak magnetism contribution are expected to contribute to nuclear decays and give non-negligible contributions to final-state correlations \cite{Cirigliano2022a}. In cases where a large cancellation of the leading order results occurs (such as for $^{19}$Ne where both $a_{\beta\nu}$ and $A_\beta$ are close to 4\%), such corrections might be sizeable. It is unclear whether the isospin breaking corrections sum coherently over all nucleons, and needs further research.

\subsubsection{Formalism mix-up and energy-dependent terms}
In parallel, care must be taken to correctly combine an experimental determination of $\rho$ with resulting calculations of the total decay rate and therefore $V_{ud}$. The importance of this was highlighted in Ref. \cite{Hayen2021}, where it was shown how recoil contributions in $\mathcal{O}(\alpha)$ calculations can appear through soft photon exchange between final states but cancel completely in the full calculation. Additionally, the calculation of such terms differs depending meant double-counting errors occurred \cite{Hayen2021}. This was due to a mismatch in the effective $\rho$ derived from experiment combined with calculations of $f_A/f_V$. Correcting this shifted the resultant $V_{ud}$ value by 2.2$\sigma$ standard deviations to bring it into agreement with neutron and superallowed determinations. 
%This was because often previous experiments extracted $\rho$ using expressions for angular correlations in the Holstein formalism, whereas calculations of $f_A/f_V$ were performed in that by Behrens-Buehring, with somewhat different different effective definitions of $\rho$.

A related question now arises with the likely appearance of energy-dependent corrections originating from the $\gamma W$ box diagram \cite{HayenIP}. Thus far, similar calculations have not been performed for the various angular correlations of the differential decay rate. If additional energy-dependent correlations were to modify these angular correlations in a way that is different from the $\beta$ spectrum, it would result in a similar mismatch in an effective $\rho$ extracted from experiment and that assumed when calculating $V_{ud}$. On the other hand, if the effect would be sizeable, this might enable a direct experimental test of the effect similar to spectral shape studies discussed in the preceding section.

\section{BEYOND STANDARD MODEL CONSTRAINTS}
\label{sec:BSM_limits}

\subsection{Cabibbo angle anomaly and exotic currents}

\begin{figure}[ht]
    \centering
        \includegraphics[width=\textwidth]{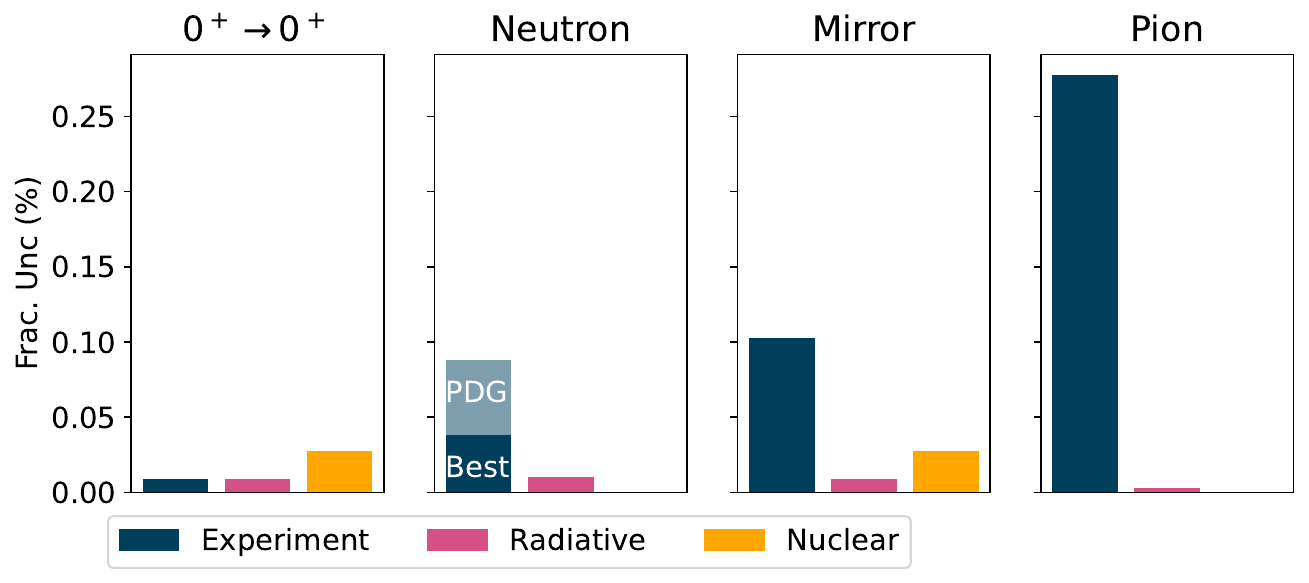}
    \caption{Fractional uncertainty in the most precise channels for a $V_{ud}$ extraction. The experimental uncertainty for the neutron is shown for the Particle Data Group average, dominated by its average for $g_A$ and a global scale factor of 2.7. Taking the most precise individual measurements leads to a total uncertainty at the 0.05\% level, to be compared to 0.03\% for $0^+\to 0^+$ decays due to nuclear structure uncertainties.}
    \label{fig:frac_unc_Vud}
\end{figure}

Together with a determination of the up-strange quark mixing matrix element one may perform a unitarity test of the top row of the CKM matrix,
\begin{equation}
    \Delta_\mathrm{CKM} = |V_{ud}|^2-|V_{us}|^2-1
    \label{eq:Delta_CKM}
\end{equation}
where contributions from $V_{ub}$ may be neglected at the current level of precision. As discussed in Sec. \ref{sec:quark_level_L}, deviations from zero in $\Delta_\mathrm{CKM}$ are sensitive to left- and right-handed currents via Eq. (\ref{eq:Vud_tilde}). Following the reanalysis of the $\gamma W$ box contribution to the radiative corrections of $\Delta_R^V$ in 2018 (Sec. \ref{sec:non_perturbative_gV}), deviations at the level of 2-3 standard deviations in Eq. (\ref{eq:Delta_CKM}) resulted in heightened scrutiny of theoretical inputs. At the same time, different ways of obtaining $V_{us}$ (from $Kl2, Kl3$ and $\tau$ decays) are internally inconsistent \cite{Falkowski2020, Cirigliano2022} despite careful examination of radiative corrections \cite{Seng2020a, Seng2019}. This situation is collectively known as the Cabibbo Angle Anomaly (CAA) \cite{Grossman2020, Coutinho2020}. Figure \ref{fig:frac_unc_Vud} shows the current status of the uncertainty budget of various channels to determine $V_{ud}$, as well as a summary of current constraints in the $V_{ud}$-$V_{us}$ plane.

\begin{figure}
    \centering
    \begin{subfigure}[b]{0.58\textwidth}
    \includegraphics[width=\textwidth]{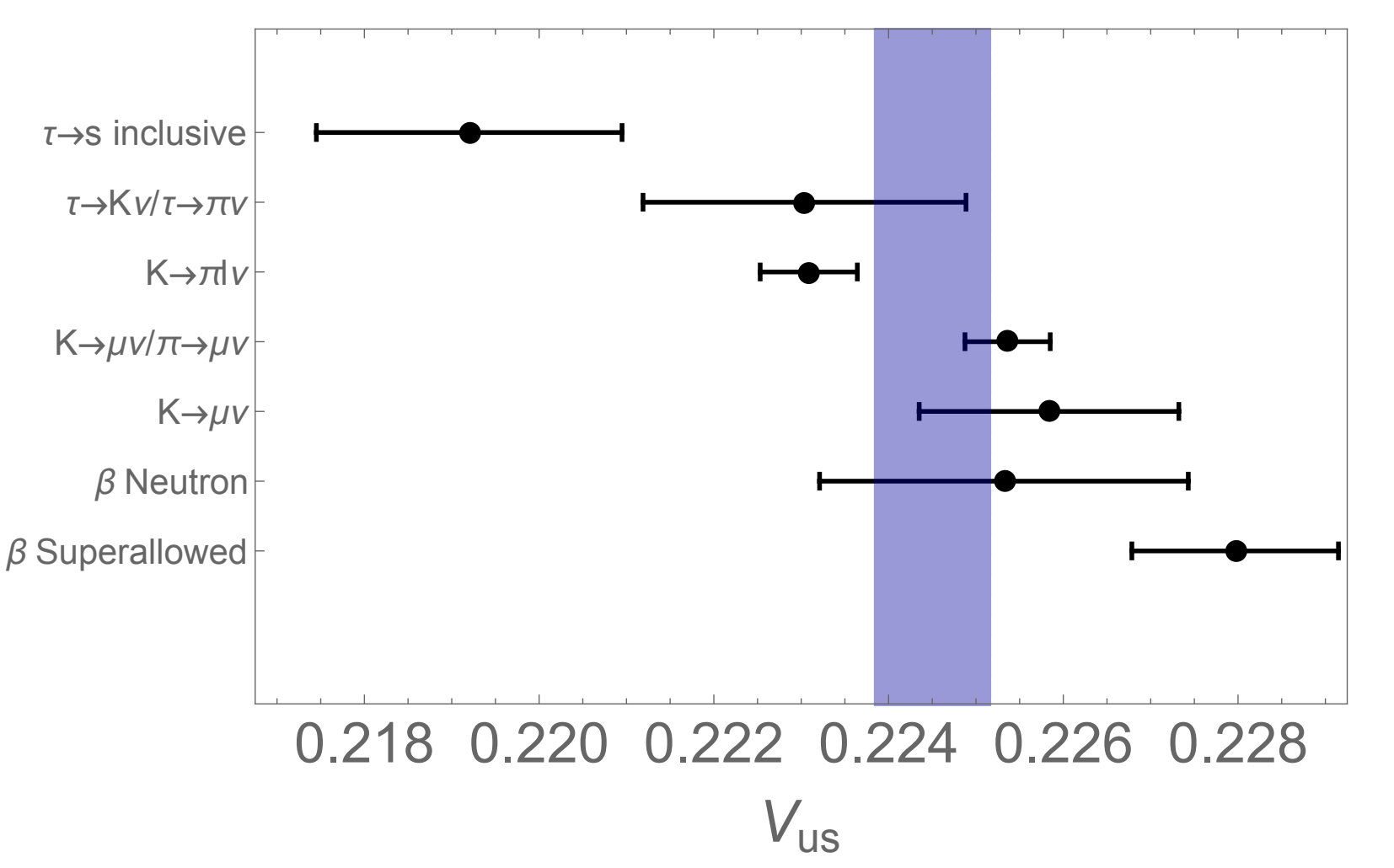}
    \end{subfigure}
    \begin{subfigure}[b]{0.38\textwidth}
        \includegraphics[width=\textwidth]{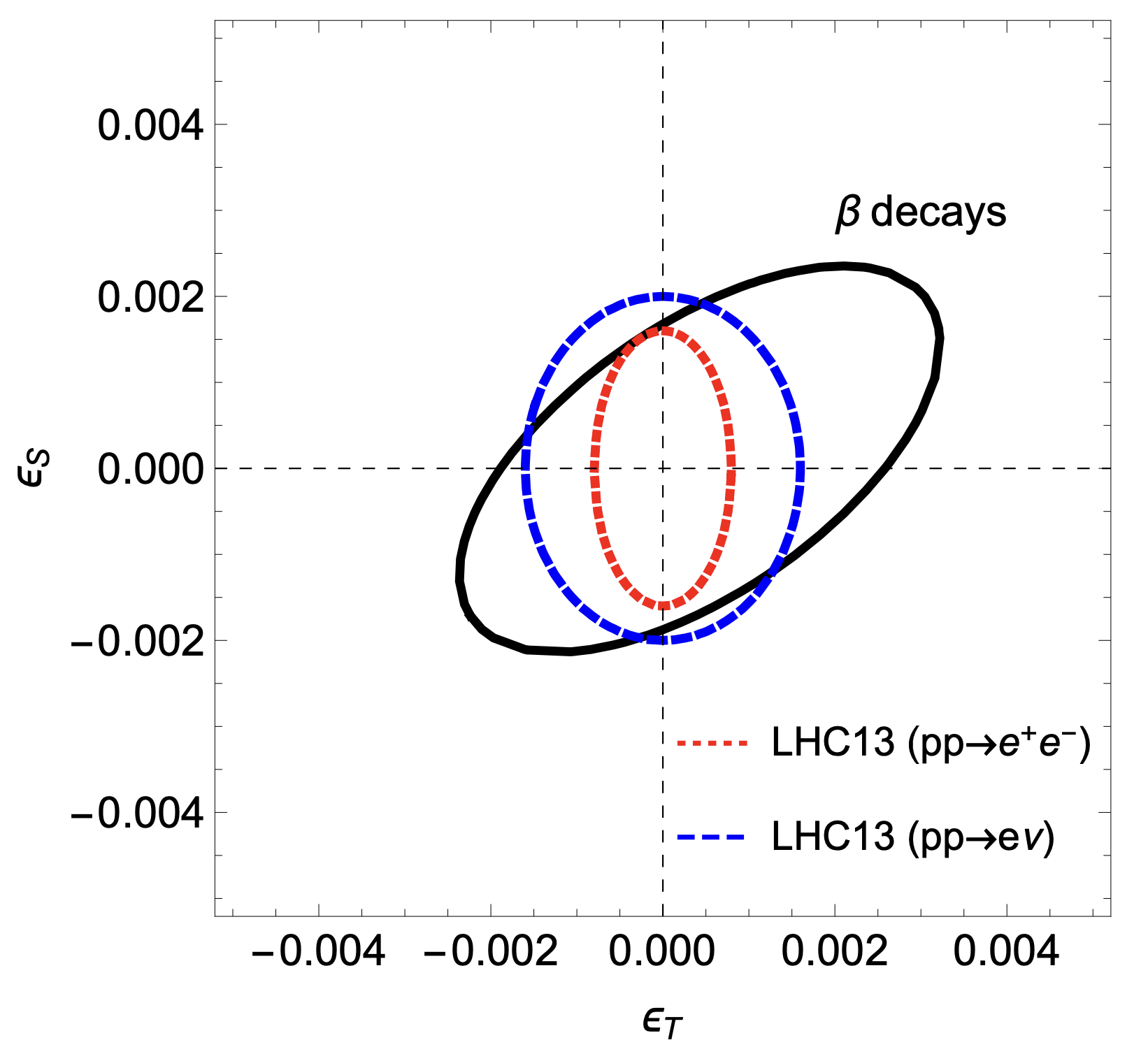}
    \end{subfigure}
    \caption{(Left) Summary of channels to determine $V_{us}$, showing the internal inconsistency. Reproduced from Ref. \cite{Cirigliano2022}. (Right) Current constraints on exotic scalar and tensor interactions with limits from $\beta$ decays and the Large Hadron Collider. Reproduced from Ref. \cite{Falkowski2020}.}
    \label{fig:summary_Vus_eps}
\end{figure}

More generally, several global analyses using the neutron and nuclear $\beta$ decay data set have been investigated \cite{Crivellin2021, Falkowski2020, Falkowski2017, Cirigliano2022, Falkowski2022, Gonzalez-Alonso2018}, with results generally consistent with Standard Model predictions, such as those shown in Fig. \ref{fig:summary_Vus_eps}. Notable exceptions include the determination of $g_A$ from the aSPECT experiment from the neutron $\beta$-$\nu$ angular correlation \cite{Beck2020, Beck2023}, and the neutron lifetime discrepancy between beam and bottle methods \cite{Wietfeldt2011, Dubbers2021}. Using a global fit \cite{Falkowski2020}, the former hints at an exotic right-handed tensor component with a significance of 3.2 $\sigma$. With the other most precise determinations of $g_A$ coming from $\beta$-asymmetry measurements \cite{Markisch2019, Brown2018, Plaster2019}, it is of great interest to see the completion of the Nab experiment at Oak Ridge National Laboratory, which aims for a determination of $g_A$ from a measurement of $a_{\beta\nu}$ at the $\mathcal{O}(3\cdot 10^{-4})$ level \cite{Pocanic2009, Broussard2019}. This provides further impetus for work at novel installations, like PERC \cite{Wang2019}.

Following reports of the anomalous $W$ boson mass measurement by the CDF collaboration \cite{Aaltonen2022}, it was additionally shown that $\beta$ decays and CKM unitarity tests play a substantial role in excluding parts of the proposed parameter space. More generally, the argument now rings very clear for a holistic analysis of all experimental measurements. Very recently, such a global analysis using collider, electroweak precision observables and low energy constraints were performed in the SMEFT \cite{Cirigliano2023b}, showing complementarity over a wide range of Wilson coefficients.

\section{CONCLUSIONS}
\label{sec:conclusions}
The study of neutron and nuclear $\beta$ decay is going through a transformation, a process which started only a number of years ago through a confluence of novel techniques and ideas, increased computational power, and tensions in the global data set. Much progress has been obtained from the theoretical side, and while substantial improvements have been obtained in, e.g., the calculation of nucleon radiative corrections and the use of lattice QCD, it will be several years before nuclear \textit{ab initio} theory will converge on calculations such as isospin symmetry breaking effects and nuclear structure effects in higher-order electroweak calculations. Nevertheless, this process is of paramount importance for the field as it promises to bring controllable theoretical uncertainties which can be systematically improved. In light of the current signs for non-unitarity of the CKM top row, its importance cannot be understated. At the same time, we have attempted to highlight several open questions resulting both from discrepancies with the traditional literature as well as unexplored contributions to several theoretical corrections.

Experimentally, a similar influx of new ideas is propelling the field forwards. Measurements using traditional scintillator or semiconductor technology have carried progress over many decades but face hard-to-avoid systematic uncertainties such as detector non-linearities, backscattering effects and bremsstrahlung losses. When these have been minimized as much as possible through clever instrumentation and set-up design, simulation approximations and their validation often dominate the systematic uncertainty budget. Nevertheless, use of proven technology allows for a quicker turnaround and will continue to provide competitive results in the foreseeable future. Novel techniques such as Cyclotron Resonance Emission Spectroscopy, however, sidestep many of these systematic uncertainty and promise to bring significant gains in precision. Similarly, the use of quantum sensors for high precision recoil spectroscopy provide yet another way of determining angular correlations to high precision. The use of Superconducting Tunnel Junctions, for example, has the benefit of fast counting rates thereby allowing on-line measurements, while the maturation of the technology was done in the synchrotron community for X-ray studies. In the future, a combination of these technologies with high efficiency atom and ion traps hold great potential to further strengthen and diversify the exotic physics reach.

\section*{DISCLOSURE STATEMENT}
The authors are not aware of any affiliations, memberships, funding, or financial holdings that might be perceived as affecting the objectivity of this review. 
%
%% Acknowledgements
\section*{ACKNOWLEDGMENTS}
The author is greatly indebted to A. Young, E. Mereghetti, W. Dekens, X. Fl\'echard and K. Leach for clarifications, inspiration and stimulating discussions. This project was supported in part by the French Agence Nationale de la Recherche.

\appendix

\section{RENORMALIZATION, RUNNING AND MATCHING}
\label{app:renormalization}
The use of EFT's is widespread and forms a cornerstone in many branches of physics \cite{Baumgart2022}. The last decade, in particular, have allowed the direct comparison of constraints from low energy up to collider searches. Following advances in electroweak EFT calculations and discrepancies in the global data set, many elements of the theoretical inputs to nuclear $\beta$ decay are being investigated using EFT methods which sometimes find conflicting answers compared to traditional calculations. There are many excellent introductions to renormalization theory, such as Refs. \cite{McComb2004, Collins1984, Ma1973}, and we will briefly summarize the procedures and introduce the jargon for experimentalists and non-experts. Specifically, we will outline the idea of renormalization group-improved perturbation theory and matching of EFTs at a common scale inspired by Refs. \cite{Buchalla1996, Jenkins2018}.

\begin{marginnote}[]
\entry{Regularization}{A way to make divergent integrals convergent by introducing a \textit{regulator}, typically a mass or length scale. Simplest examples include enforcing integration limits, such as a non-zero photon mass at low momenta or an artificial high energy limit of applicability.}
\entry{Renormalization}{Following regularization, a redefinition of a \textit{finite} number of free parameters in the Lagrangian such that measurable quantities are independent of the renormalization scale.}
\end{marginnote}
Even the most basic quantum field theories contain infinities starting at low number of loops, running into divergences at low (infrared, IR) and high (ultraviolet, UV) momentum scales. Unsurprisingly, the Standard Model contains divergences beyond tree-level in $d=4$ space-time dimensions, which can be regularized using various renormalization schemes. The latter corresponds to the introduction of a UV cutoff scale to separate finite from divergent parts of the integral, and since physical observables must be independent of this scale, the separation is necessarily arbitrary. The most popular device in perturbative calculations is dimensional regularization, where the number of space-time dimensions $d$ is taken to be continuous such that the integral converges (decreasing $d$ for UV divergences and increasing for IR divergences). The result is afterwards analytically continued to $d=4$ picking up poles in divergent quantities. Similarly to the introduction of an integration cut-off, a mass scale $\mu$ occurs, e.g., through a redefinition of a bare (unrenormalized) coupling constant, $g_0$, such that it remains dimensionless, i.e. $g_0 = g\mu^\epsilon$ for $d=4-2\epsilon$. We may then write integrals that would, e.g., diverge logarithmically for $d=4$,
\begin{equation}
    -ig_0^2\mu^{2\epsilon} \int \frac{d^dq}{(2\pi)^d} \frac{1}{(q^2-m^2)^2} = \frac{g_0^2}{16\pi^2}\left[\frac{1}{\epsilon}-\gamma_E+\ln{4\pi}-\ln(m^2/\mu^2)\right]
    \label{eq:dimreg_example_int}
\end{equation}
\begin{marginnote}[]
\entry{Wick rotation}{Multiplying the time-component of a four-dimensional integration constant by $i$, transforming the vector product from the Minkowski to a Euclidean metric, used to more easily solve integrals.}
\end{marginnote}
as $\int d^4q/q^4 \to \infty$. The technical steps to reach the result of Eq. (\ref{eq:dimreg_example_int}) include performing a Wick rotation and using properties of the Beta Euler function, but are not our main interest here. The $d=4$ divergence is contained in the $1/\epsilon$ term, but the final result also includes surviving finite terms, some depending on $\mu$. Many - an infinite number, in fact - diagrams in the same theory will give rise to similarly diverging integrals. In renormalizable theories, one is able to ` subtract' the poles by adding to the Lagrangian a \textit{finite} number of `counterterms' such that the full theory is independent of $\epsilon$ and $\mu$ to \textit{infinite} order in the perturbation expansion\footnote{With the advent of effective field theories, the philosophy on renormalizability as a criterion for a useful theory has shifted, requiring instead a finite number of a counterterms \textit{order-by-order} in the theory.}. This procedure is performed through what is called a \textit{renormalization scheme}, of which the final result for measurable quantities (cross sections, decay rates, etc.) is necessarily independent. In the (modified) minimal subtraction scheme, MS ($\overline{\mathrm{MS}}$), one subtracts only the $1/\epsilon$ $(1/\epsilon-\gamma_E+\ln{4\pi})$ pole terms leaving the characteristic logarithmic dependence on the ratio of mass, $m$, and renormalization scale, $\mu_\mathrm{MS}$, as the surviving finite piece. This implies that masses, couplings, etc., depend on the renormalization scale by construction.
%\footnote{Due to presence of inherently 4-dimensional objects such as the $\gamma^5$ matrix or the Levi-Civita tensor when dealing with, e.g., left-handed fermions in weak interactions, different schemes are introduced to deal with these complexities, such as Naive Dimensional Regularization (NDR) or 't Hooft-Veltman.}

Calculations to higher-order calculations in the perturbation expansion will contain powers of $\ln^n(m^2/\mu^2)$ so that its convergence depends on these contributions being small. One may choose $\mu\sim\mathcal{O}(m)$ rendering logarithmic corrections small, at the price of defining the parameter (coupling, mass, $\ldots$) \textit{at that mass scale} which often does not correspond to what is measured experimentally. A way out between these conflicting situations (poor convergence in the perturbation expansion, on the one hand, and parameters not defined at the scale experiments probe it, on the other), is offered by the renormalization group which specifies how parameters change as we change the renormalization scale. This will allow us to smoothly connect different scales, an essential ingredient in matching different effective field theories.

Constructing the Lagrangian of an effective field theory as a low-energy limit (at a scale $m$) of a UV theory (at a scale $M$) consists of choosing the relevant degrees of freedom and writing down all possible terms consistent with the UV gauge symmetries, i.e.
\begin{equation}
    \mathcal{L}^M = \mathcal{L}_{free}^M + \mathcal{L}_{int}^M \qquad \mathrm{and} \qquad \mathcal{L}_{eff}^m = \mathcal{L}_{free}^m + \sum_{i}C_i O_i
\end{equation}
where $\mathcal{L}_{free}$ is the free-field Lagrangian, $\mathcal{L}_{int}$ contains interactions terms and $C_i$ are Wilson coefficients for operators $O_i$ constructed from the EFT fields at scale $m$. If the UV theory is known, one may try to determine the Wilson coefficients through a \textit{matching} procedure, i.e. calculating all possible diagrams to the required order in both theories with light external fields and comparing terms. This typically results in Wilson coefficients with 'large logarithms', i.e. terms that contain $\ln{M^2/\mu^2}$. Since the operators $O_i$ (and their Wilson coefficients $C_i$) are calculated at a scale $\mu\sim m \ll M$, these large logarithms cause the perturbation theory to have poor convergence as matching results in terms containing $\log^n(M^2/m^2)$. A way to remedy this situation is using the renormalization group, which specifies how renormalized couplings, masses, etc., change under a renormalization scale change such that the final result remains independent of $\mu$. This information is encoded in a renormalization group equation (RGE), a differential equation for each parameter of the theory. A common example is the $\beta$ function, which describes the change in the effective coupling as a function of the renormalization scale
\begin{equation}
    \frac{d}{d~\ln \mu} g(\mu) = \beta(g(\mu)).
    \label{eq:beta_function}
\end{equation}
Equation (\ref{eq:beta_function}) determines the `running of the coupling', and $\beta$ can be obtained, order-by-order, by doing loop diagram calculations\footnote{Theories where $\beta > 0$, such as QED, are perturbative at low energies, whereas for $\beta < 0$, such as in QCD, effects become perturbative at high energies and imply asymptotic freedom.}. Similarly, one may construct an RGE for each Wilson coefficient, $C_i$, where the terms are known as \textit{anomalous dimensions}, $\gamma$, which typically depend on the running couplings in the theory. Schematically, we may write
\begin{equation}
    \frac{d}{d~\ln{\mu}}\alpha = -2\beta_0 \frac{\alpha^2}{4\pi} + \mathcal{O}(\alpha^3) \qquad \frac{d}{d~\ln{\mu}}C_i(\mu) = \left[\frac{\alpha}{4\pi}\gamma \right]C_i(\mu) 
\end{equation}
where $\alpha = g^2/4\pi$. This system of equations may be solved to give
\begin{equation}
    C_i(\mu) = \left[\frac{\alpha(M)}{\alpha(\mu)}\right]^{\gamma/2\beta_0}C_i(M),
\end{equation}
where it is now clear how the anomalous dimension non-trivially affects the scaling due to quantum effects. 

Putting everything together, the use of the RGE allows one to improve the naive perturbation theory result, as one may then consider the renormalization scale change from the UV theory at $\mu\sim M$ to the EFT at $\mu \sim m$ through a series of infinitesimal changes connected via the RGE. If $\gamma$ is known to leading order, we are including the large-logarithm terms to all orders in perturbation theory \textit{automatically}, i.e. $[\alpha \ln(M^2/\mu^2)]^n$ while including $\gamma$ at next-to-leading-order sums terms of order $\alpha [\alpha \ln(M^2/\mu^2)]^n$. In conclusion, whereas naively connecting theories defined at two different scales inherently introduces what is known colloquially as 'large-logarithms', using renormalization groups is a way to `softly' connect two different energy regimes by smoothly going from one mass scale to the other, integrating incremental changes along the way while also retaining good convergence properties.

\section{RECOIL-ORDER MATRIX ELEMENTS}
\label{app:recoil_order}
When the energy release in a decay is much smaller than the mass of the initial and final states small recoil corrections appear, i.e. contributions of $\mathcal{O}(q/M) \ll 1$, where $q$ is the momentum transfer during the decay and $M$ in the mass of the decaying particle. For $\beta$ decay this energy release almost never exceeds 10 MeV, meaning $q/M \sim 10^{-3}$. At the current level of experimental precision, however, these terms are relevant. This fact is exacerbated when significant cancellations occur in the main matrix elements, so that these recoil-order effects are significantly boosted in relative precision \cite{Hayen2020a}. Following Holstein \cite{Holstein2018}, it is useful to categorize recoil-order terms following their origin
\begin{itemize}
    \item Kinematic, $ \mathcal{O}(1) \times q/M$
    \item Dynamic, $\mathcal{O}(A) \times q/M$
    \item Coulombic, $\mathcal{O}(\alpha Z MR) \times q/M$
\end{itemize}
where $A$ is the mass number of the decaying nucleus and $R$ is the charge radius. Points two and three are contained in the proper description of the transition matrix element. 

When discussing recoil-order matrix elements one is typically concerned with the second category. This can be understood simply from the nucleon system, where the Lorentz invariant decomposition into form factors of the bilinear quark operators in Sec. \ref{sec:formalism} may be written as follows
\begin{eqnarray}
    \langle p | \bar{u}\gamma^\mu d | n \rangle &=& \bar{u}_p \left[g_V(q^2)\gamma^\mu + \frac{\tilde{g}_M(q^2)}{2m_N}\sigma^{\mu\nu}q_\nu + \frac{\tilde{g}_S(q^2)}{2m_n}q^\mu\right]u_n \label{eq:pn_vector}\\
    \langle p | \bar{u}\gamma^\mu \gamma^5d | n \rangle &=& \bar{u}_p \left[g_A(q^2)\gamma^\mu + \frac{\tilde{g}_T(q^2)}{2m_N}\sigma^{\mu\nu}q_\nu + \frac{\tilde{g}_P(q^2)}{2m_n}q^\mu\right]\gamma^5u_n \label{eq:pn_axial_vector} \\
    \langle p | \bar{u} d | n \rangle &=& g_S(0)\bar{u}_p u_n + \mathcal{O}(q^2/m_n^2)\\
    \langle p | \bar{u}\gamma^5 d | n \rangle &=& g_P(0)\bar{u}_p \gamma^5 u_n + \mathcal{O}(q^2/m_n^2)\\
    \langle p | \bar{u} \sigma^{\mu\nu} d | n \rangle &=& g_T(0)\bar{u}_p \sigma^{\mu\nu} u_n + \mathcal{O}(q^2/m_n^2)
\end{eqnarray}
where $q = p_n-p_p$ and $m_N$ is the nucleon mass in the isospin limit, and where we neglected additional momentum dependence of the exotic (non-($V-A$)) interactions. Here, $\tilde{g}_{M, S, T, P}$ are known as the weak magnetism, induced scalar, induced tensor, and induced pseudoscalar, respectively. Of all of these, the weak magnetism contribution gives rise to percent-level corrections to the $\beta$ spectrum (with $\tilde{g}_M = 4.706$), whereas others are zero by symmetry ($\tilde{g}_{S, T}$) or give only very small contributions ($\tilde{g}_P$). This form factor decomposition generalizes to nuclear systems with higher spins as discussed in Sec. \ref{sec:nuclear_level_L}, and is discussed in more detail in Refs. \cite{Holstein1974, Hayen2018}.

\bibliographystyle{ar-style5.bst}
\bibliography{library}

\end{document}